\documentclass[showpacs,preprint,preprintnumbers
,nofootinbib,prd,
  eqsecnum,10pt
]{revtex4-1}
  
\usepackage{amsmath,amssymb,amsfonts}
\usepackage{mathrsfs}
\usepackage[normalem]{ulem}
\usepackage{textcomp}
\usepackage{hyperref}
\usepackage{bm}
\usepackage[bottom]{footmisc}
\usepackage{tensor}
\usepackage{graphicx}
\usepackage{psfrag}
\usepackage[usenames,dvipsnames]{xcolor}
\usepackage[utf8]{inputenc}
\usepackage[english, capitalise]{cleveref}

\crefname{chapter}{Chap.}{Chap.}
\crefname{section}{Sec.}{Sec.}
\Crefname{chapter}{Chapter}{Chapters}
\Crefname{section}{Section}{Sections}
\Crefname{eqs}{Eqs.}{Eqs.}
\Crefformat{eqs}{Eqs.~(#2#1#3)}

\newcommand{\tet}[1]{\hat{#1}}

\newcommand{\dd}{\mathrm{d}}
\newcommand{\di}{\mathrm{i}}
\newcommand{\de}{\mathrm{e}}

\definecolor{darkgreen}{rgb}{0,0.5,0}

\hypersetup{
    bookmarks=true,
    unicode=false,
    pdftoolbar=true,
    pdfmenubar=true,
    pdffitwindow=false,
    pdfstartview={FitH},
    pdftitle={My title},
    pdfauthor={Author},
    pdfsubject={Subject},
    pdfcreator={Creator},
    pdfproducer={Producer},
    pdfkeywords={keyword1} {key2} {key3},
    pdfnewwindow=true,
    colorlinks=true,
    linkcolor=red,
    citecolor=cyan,
    filecolor=magenta,
    urlcolor=darkgreen,
    linktocpage=true
  }

\allowdisplaybreaks

\begin{document}

\title{Tidal effects in the equations of motion of compact binary systems \\
to next-to-next-to-leading post-Newtonian order}

\author{Quentin \textsc{Henry}}\email{henry@iap.fr}
\affiliation{$\mathcal{G}\mathbb{R}\varepsilon{\mathbb{C}}\mathcal{O}$, 
Institut d'Astrophysique de Paris,\\ UMR 7095, CNRS, Sorbonne Universit{\'e},\\
98\textsuperscript{bis} boulevard Arago, 75014 Paris, France}

\author{Guillaume \textsc{Faye}}\email{faye@iap.fr}
\affiliation{$\mathcal{G}\mathbb{R}\varepsilon{\mathbb{C}}\mathcal{O}$, 
Institut d'Astrophysique de Paris,\\ UMR 7095, CNRS, Sorbonne Universit{\'e},\\
98\textsuperscript{bis} boulevard Arago, 75014 Paris, France}

\author{Luc \textsc{Blanchet}}\email{blanchet@iap.fr}
\affiliation{$\mathcal{G}\mathbb{R}\varepsilon{\mathbb{C}}\mathcal{O}$, 
Institut d'Astrophysique de Paris,\\ UMR 7095, CNRS, Sorbonne Universit{\'e},\\
98\textsuperscript{bis} boulevard Arago, 75014 Paris, France}
 
\date{\today}

\begin{abstract}
  As a first step in the computation of the orbital phase evolution of
  spinless compact binaries including tidal effects up to the
  next-to-next-to-leading (NNL) order, we obtain the equations of
  motion of those systems and the associated conserved integrals in
  harmonic coordinates. The internal structure and finite size effects
  of the compact objects are described by means of an effective
  Fokker-type action. Our results, complete to the NNL order,
  correspond to the second-post-Newtonian (2PN) approximation beyond
  the leading tidal effect itself, already occurring at the 5PN
  order. They are parametrized by three polarizability (or
  deformability) coefficients describing the mass quadrupolar, mass
  octupolar and current quadrupolar deformations of the objects
  through tidal interactions. Up to the next-to-leading (NL) order, we
  recover previous results in the literature; up to the NNL order for
  quasi-circular orbits, we confirm the known tidal effects in the (PN
  re-expansion of the) effective-one-body (EOB) Hamiltonian. In a
  future work, we shall derive the tidal contributions to the
  gravitational-wave flux up to the NNL order, which is the second
  step required to find the orbital phase evolution.
\end{abstract}

\maketitle

\section{Introduction}\label{sec:introduction}

The direct detection of gravitational waves (GW) generated by the
orbital motion and merger of compact binary systems~\cite{GW150914,
  GW170817} opens up a new avenue in fundamental physics. Notably, it
will play a paramount role in understanding the physics of compact
objects, mainly black holes or neutron stars. The tidal effects
between such objects are particularly interesting because they permit
revealing and probing their internal structure, as well as eventually
distinguishing between black holes, neutron stars or, possibly, more
exotic entities like boson stars~\cite{FaberRasioLR, BuonSathya15}.

The tidal interaction affects both the conservative equations of
motion (EoM) and the GW emission of the compact binary system. This
results in a modification of the time evolution of the binary's
orbital frequency and phase which is directly observable (see
\textit{e.g.}~\cite{MW03, FH08, DNV12, F14}).  The tidal distortion
depends on the Love numbers~\cite{Love11}, characterizing the rigidity
and the deformability of the body, \textit{i.e.}  its capacity to
change shape under the influence of an external tidal field.  Those
Love numbers depend in turn on the internal equation of state (EoS) of
the body, which is uncertain at high densities~\cite{Hind08,
  HindLLR10}. They decrease as the compactness of the body increases,
reaching zero in the limit of a maximally compact object,
\textit{i.e.}, for a black hole~\cite{FangLove05, BinnP09, DN09tidal}.

The leading tidal contributions to the orbital dynamics are due to
quadrupolar deformations and, for compact binaries, manifest
themselves as formally very small corrections in the accelerations, of
the order of 5PN or $\sim (v/c)^{10}$, where $v$ denotes the relative
orbital velocity. However, the 5PN coefficient appearing in front of
the small 5PN factor $(v/c)^{10}$ can be quite large and the effect is
measurable.\footnote{One can speculate that the tidal 5PN coefficient
  is larger than the purely orbital 5PN contribution to the orbital
  phase for point particles, which is currently unknown.} It scales
like the dimensionless parameter
\begin{equation}\label{eq:Lambda}
\Lambda^{(2)} = \frac{2}{3}
k^{(2)} \biggl(\frac{R c^2}{G m}\biggr)^5\,,
\end{equation}
where $k^{(2)}$ denotes the mass-type quadrupolar second Love number
of the body, while $m$ and $R$ represent its mass and radius.
Typically, the compactness parameter $\mathcal{C}\sim G m/(R c^2)$ is
of order $0.15$ for neutron stars while the Love number is
$k^{(2)}\sim 0.1$ (depending on the EoS)~\cite{BinnP09, DN09tidal},
hence we expect $\Lambda^{(2)}\sim 1000$. With the binary neutron star
event GW170817~\cite{GW170817}, the detectors LIGO and Virgo have
already been able to put an observational constraint on the
particular combination of $\Lambda^{(2)}_1$, $\Lambda^{(2)}_2$ and the
masses that enter the orbital phase evolution of the two neutron
stars~\cite{FH08, F14}. This constraint permitted excluding some of
the stiffest EoS, for which the neutron stars are less
compact~\cite{LVCpropertiesGW170817, LVCO1O2}. However, the majority
of softer EoS are still allowed (see also~\cite{LVCcomparisonGW170817}
and references therein).

The problem of tidal interactions between compact objects beyond the leading
quadrupolar level has been addressed in Refs.~\cite{VHF11, BiniDF12, VF13,
  AGP18, BV18, Landry18}. The conservative dynamics, from which follow the
EoM, was obtained in the work~\cite{AGP18} at leading order but including
linear spin couplings, or in~\cite{VF13} and~\cite{BiniDF12} up to the
next-to-leading (NL) and the next-to-next-to-leading (NNL) orders,
respectively, while the energy flux, waveform amplitude and phase evolution
have been computed to the leading order in the presence of spin couplings, and
NL order, equivalent to the formal 6PN level~\cite{VHF11, BV18, Landry18},
in the non-spinning case.\footnote{The NNL order in the dynamics corresponds
  to 2PN order beyond the leading 5PN quadrupolar tidal effect and is thus
  formally equivalent to a 7PN orbital effect; similarly, the NL order means
  1PN beyond the leading 5PN effect.} The tidal interactions in both the
dynamics and waveform have also been included in the effective-one-body (EOB)
models for template generation~\cite{DNV12, BiniDF12}.

In the present paper, we compute the tidal effects in the
conservative EoM, as well as all associated conserved quantities, at
the NNL order for spinless neutron stars on generic binary orbits in
harmonic coordinates. We follow closely the method proposed in
Ref.~\cite{BiniDF12}, describing the internal structure and finite
size effects of the compact objects by means of an effective
Fokker-type action. Our final NNL results are parametrized by three
polarizability (or deformability) coefficients describing the mass
quadrupolar, mass octupolar and current quadrupolar deformations of
the objects through tidal interactions. In the case of quasi-circular
orbits, we confirm the expression of the tidal terms in the EOB
Hamiltonian up to the NNL order~\cite{BiniDF12}. So as to compute the
tidal contribution to the orbital phase at the NNL order, we 
need both the conservative NNL energy of the system and the GW energy
flux at the same NNL order. In a forthcoming
paper~\cite{article_flux}, we shall complete the present work by
computing the latter effect for the GW flux, which will yield the
orbital phase evolution at the NNL order.

Although the knowledge of the NNL/2PN relative tidal effect is
probably not directly useful for the data analysis of the advanced
LIGO and Virgo detectors, it may become relevant for the future
third-generation detectors, like the Einstein Telescope or the Cosmic
Observatory. On the other hand, detailed comparisons with numerical
relativity (NR) simulations of binary neutron-star mergers require the
control of high-order tidal interactions on the analytic side. Yet,
such comparisons are essential to get a grip on the errors of the
predicted waveforms and to properly calibrate EOB models. More
generally, adding analytic tidal effects on the top of PN templates of
point particles is a good way of controlling the systematic errors due
to our lack of knowledge of the higher-order terms in the PN
expansion~\cite{F14, BuonSathya15}.

This article is organized as follows. In~\cref{sec:action}, we define
the effective Fokker action with appropriate non-minimal matter
couplings describing finite size effects. The quantities entering this
action are determined by the 2PN metric, presented
in~\cref{sec:metric} and computed off-shell, \textit{i.e.}, without
replacement of accelerations by the EoM, ready for insertion into the
action. Our final Lagrangian, accurate to NNL leading order for tidal
effects, is displayed in~\cref{sec:Fokker}, together with the
associated NL center-of-mass (CoM) position. We then derive,
in~\cref{sec:CoM}, the tidal dynamics in the CoM frame for general
orbits, as well as the reduction for quasi-circular orbits.
The~\cref{appendix:Newtonian} is devoted to basic recalls and
motivation concerning the treatment of tidal effects in the Newtonian
theory.  In~\cref{appendix:proof} we show, using standard techniques
of Lagrangian formalism, that the tidal multipole moments up to the
NNL order can be defined equivalently by means of either the Riemann
tensor or the Weyl tensor. Finally we give in~\cref{appendix:accNNL}
the complete tidal acceleration in a general frame for arbitrary
orbits to NNL order.

\section{Effective Fokker action with non-minimal matter
  couplings} \label{sec:action}

The model we use is defined by the gravitation-plus-matter action
$S=S_g+S_m$, where the gravitational part $S_g$ is the standard
Einstein-Hilbert action, to which we add the appropriate
harmonic-gauge fixing term:
\begin{equation}\label{eq:Sg}
S_{g} = \frac{c^{3}}{16\pi G} \int \dd^{4}x \, \sqrt{-g} \left[
  R -\frac{1}{2}
  g_{\mu\nu}\Gamma^{\mu}\Gamma^{\nu} \right] \,,
\end{equation}
where $R$ is the curvature scalar, $\Gamma^{\mu}_{\rho\sigma}$ is the
usual Christoffel symbol, and $\Gamma^{\mu} =
g^{\rho\sigma}\Gamma^{\mu}_{\rho\sigma}$. In practical calculations,
we rather use the Laudau-Lifshitz~\cite{LL} form of the
action.\footnote{Throughout the paper, we use the conventions of
  MTW~\cite{MTW}; in particular, the metric signature is $(-,+,+,+)$
  and the Riemann tensor satisfies the identity
  $(\nabla_{\mu}\nabla_{\nu} - \nabla_{\nu}\nabla_{\mu})V^{\lambda} =
  R\indices{^\lambda_\kappa_\mu_\nu}V^{\kappa}$.}

The matter part of the action $S_m$ describes massive point-like
particles with internal structure. It contains specific non-minimal
couplings to the space-time curvature that describe the finite size
effects of the compact bodies solely due to the tidal interactions,
all spins being taken to zero. Since the matter action is regarded
as localized on the worldline of the particles, it is generally
referred to as a ``skeletonized'' effective action.  In order to
define it, we introduce a local inertial coordinate frame along each
body worldline, together with the associated local tetrad
$e_{\tet{\alpha}}^{\phantom{\tet{\alpha}}\mu}$. More precisely, we
pose $e_{\tet{\alpha}}^{ \phantom{\tet{\alpha}}\mu} = \partial
x^\mu/\partial X^{\tet{\alpha}}$, where $\{x^\mu\}$ is a global
coordinate system and $\{X^{\tet{\alpha}}\}$ is the local inertial
frame in the vicinity of the body in question. We may choose
$\{X^{\tet{\alpha}}\}$ to be a Fermi local normal coordinate
system~\cite{Fermi22, MM63}, so that the tetrad is orthonormal on the
worldline, the time coordinate of the Fermi coordinates coincides with
the proper time along the worldline, and the zero-th time-like tetrad
vector is the four velocity of the particle. In its own local frame,
the body feels the tidal multipole moments generated by the other
bodies at its very location, namely the $\ell$-th order mass-type
moments $G_{\tet{L}}$ and the current-type ones $H_{\tet{L}}$, where
those quantities refer to the spatial tetradic components of the
moments, \textit{i.e.} projected along the local tetrad, with
$\tet{L}=\tet{i}_1\cdots \tet{i}_\ell$ denoting a multi-spatial index
composed of $\ell$ spatial tetradic indices.

In this paper, we assume that each body stays in static
equilibrium at any instant. In the absence of spin, the internal
structure is then entirely determined by the mass and the EoS. Thus,
the elementary bricks that are allowed to construct $S_m$ are tensors
defined from the metric only and evaluated at the given particle
position, with all indices contracted so as to preserve the invariance
under rotation and parity in the corresponding constant-time
hypersurface of the local Fermi rest frame. For our purpose, it will be
sufficient to consider the same non-minimal terms as in
Ref.~\cite{DN10}, built from quadratic (kinetic-like) couplings in the
tidal moments $G_{\tet{L}}$ and $H_{\tet{L}}$. Hence the form of the
matter action (adding also the particle's label
$A\in\{1,2\}$)\footnote{The constant mass of body $A$ is denoted $m_A$
  and its proper time $\dd\tau_{A} = (-[g_{\mu \nu}]_A\dd
  y^{\mu}_{A}\dd y^{\nu}_{A}/c^2)^{1/2}$, where $y_A^\mu(\tau_A)$ is
  the particle's worldline. The four velocity $u^{\mu}_{A} = \dd
  y^{\mu}_{A}/(c\,\dd\tau_{A})$ is such that $[g_{\mu\nu}]_A
  u_{A}^{\mu}u_{A}^{\mu} = -1$, with $[g_{\mu \nu}]_A$ denoting the
  metric regularized at the location of body $A$; this is of course
  nothing but the time-time component of the orthonormalizing
  condition of the tetrad, $\eta_{\hat{\alpha}\hat{\beta}} =
  [g_{\mu\nu}]_A\,e_{\tet{\alpha}}^{A\mu}\,e_{\tet{\beta}}^{A\nu}$.}
\begin{equation}\label{eq:Sm}
S_{m} = \sum_{A} \int \dd \tau_{A} \left\{ - m_A c^2 + \sum_{\ell=2}^{+\infty}
  \frac{1}{2\ell!} \biggl[ \mu_{A}^{(\ell)}
  \, \bigl(G^{A}_{\hat{L}}\bigr)^2 +
  \frac{\ell}{(\ell+1)c^{2}} \,\sigma_{A}^{(\ell)}
  \bigl(H^{A}_{\tet{L}}\bigr)^2 \biggr]  + \cdots \right\}\,.
\end{equation}
The ellipsis indicate many higher-order non-linear combinations of the tidal
moments and their covariant (proper-time) derivatives, which we do not need to
include here (see \textit{e.g.} Eq.~(2.3) of~\cite{BiniDF12}). For more
insight and motivation about the non-minimal action, see Refs.~\cite{ThH85,
  Zhang86, DSX1, BiniDF12} and the treatment of tidal effects in the Newtonian
model as recalled in~\cref{appendix:Newtonian}.

The above tidal moments are given by appropriate covariant derivatives of the
Weyl tensor. We define first the spatial tetradic components of the moments
appearing in~\cref{eq:Sm} (for $\ell\geqslant 2$) as
\begin{subequations}\label{eq:defGH}
\begin{align}
G^A_{\tet{L}} &= - c^2
\Bigl[\nabla_{\langle\tet{i}_1}\cdots\nabla_{\tet{i}_{\ell-2}}
  C_{\tet{i}_{\ell-1}\underline{\tet{0}}\tet{i}_\ell\rangle\tet{0}}\Bigr]_A
\,,\\ H^A_{\tet{L}} &= 2 c^3
\Bigl[\nabla_{\langle\tet{i}_1}\cdots\nabla_{\tet{i}_{\ell-2}}
  \,C^{*}_{\tet{i}_{\ell-1}\underline{\tet{0}}\tet{i}_\ell\rangle\tet{0}}\Bigr]_A\,.
\end{align}
\end{subequations}
The angle brackets over the $\ell$ free spatial indices $\tet{L} =
\tet{i}_1\cdots \tet{i}_\ell$ of the above tensor expressions means
that they must be replaced by their symmetric and trace-free (STF)
parts over those indices, the underlined indices being excluded from
the STF projection. We denote by $\nabla_{\tet{\alpha}}$ the usual
covariant tetradic derivative [we pose
  $\tet{\alpha}=(\tet{0},\tet{i})$], whereas
$C_{\tet{\alpha}\tet{\beta}\tet{\gamma}\tet{\delta}}$ and
$C^{*}_{\tet{\alpha}\tet{\beta}\tet{\gamma}\tet{\delta}}$ represent
the tetradic components of the Weyl tensor (whose definition is
recalled in \cref{eq:Weyl} below) and its dual.\footnote{In our
  convention, $C^{*}_{\tet{\alpha}\tet{\beta}\tet{\gamma}\tet{\delta}}
  \equiv \frac{1}{2}
  \varepsilon_{\tet{\alpha}\tet{\beta}\tet{\eta}\tet{\zeta}}
  \,C^{\tet{\eta}\tet{\zeta}}_{\phantom{\tet{\eta}\tet{\zeta}}\tet{\gamma}\tet{\delta}}$
  or, in covariant form, $C^{*}_{\mu \nu \rho \sigma} \equiv
  \tfrac{1}{2}\varepsilon_{\mu \nu \lambda \kappa}
  \,C\indices{^\lambda^\kappa_\rho_\sigma}$, where
  $\varepsilon_{\tet{\alpha}\tet{\beta}\tet{\gamma}\tet{\delta}}$
  denotes the tetradic components of the completely anti-symmetric
  Levi-Civita tensor $\varepsilon_{\mu\nu\rho\sigma}$, defined by
  $\varepsilon_{\tet{0}\tet{1}\tet{2}\tet{3}}=1$ and
  $\varepsilon_{0123}=\sqrt{-g}$. The tetradic covariant derivative
  obeys, \textit{e.g.}, $\nabla_{\tet{\alpha}}V^{\tet{\beta}} =
  e_{\tet{\alpha}}^{\phantom{\tet{\alpha}}\mu}
  e^{\tet{\beta}}_{\phantom{\tet{\beta}}\nu}\nabla_{\mu}V^{\nu}$.}  By
construction, the tidal moments~\eqref{eq:defGH} are symmetric over
their spatial indices $\tet{L}$ and all their traces are zero,
\textit{i.e.},
$\delta_{\tet{i}_1\tet{i}_2}G_{\tet{i}_1\tet{i}_2\cdots\tet{i}_\ell}=0$.

Next, we introduce the covariant versions of the previous tidal
tensors. Since $u^\mu=e_{\tet{0}}^{\phantom{\tet{0}}\mu}$, this is
achieved by imposing that they live in the particle's local spatial
hypersurface, which is orthogonal to the four velocity. Thus, we
complete the definition of the tidal moments~\eqref{eq:defGH} by
requiring them to obey
\begin{equation}\label{eq:G0i}
  G^A_{\tet{0}\tet{\alpha}_2\cdots\tet{\alpha}_\ell} =
  H^A_{\tet{0}\tet{\alpha}_2\cdots\tet{\alpha}_\ell}=0\,.
\end{equation}
In this way, $G_{\tet{\alpha}_1\cdots\tet{\alpha}_\ell}$ and
$H_{\tet{\alpha}_1\cdots\tet{\alpha}_\ell}$ are both Lorentz tensors
and covariant scalars, while their covariant versions in an arbitrary
coordinate system $\{x^\mu\}$ read
\begin{subequations}\label{eq:defGHcov}
\begin{align}
G^A_{\mu_1\cdots\mu_\ell} &= - c^2 \Bigl[\nabla^\perp_{\langle
    \mu_1}\cdots\nabla^\perp_{\mu_{\ell-2}}
  C_{\mu_{\ell-1}\underline{\rho}\mu_\ell\rangle\sigma}\Bigr]_A
u_A^\rho \,u_A^\sigma \,,\\ H^A_{\mu_1\cdots\mu_\ell} &= 2 c^3
\Bigl[\nabla^\perp_{\langle \mu_1}\cdots\nabla^\perp_{\mu_{\ell-2}}
  \,C^{*}_{\mu_{\ell-1}\underline{\rho}\mu_{\ell}\rangle\sigma}\Bigr]_A
u_A^\rho \,u_A^\sigma \,.
\end{align}
\end{subequations}
Here, we denote $\nabla^\perp_\mu=\perp_\mu^\nu \nabla_\nu$, with
$\perp_\mu^\nu = \delta_\mu^\nu + u_\mu u^\nu$ being the projector onto the
hypersurface orthogonal to the four velocity [notice that
  $\perp_{\tet{\alpha}}^\mu = (0,
  e_{\tet{i}}^{\phantom{\tet{i}}\mu})$]. The tidal moments are both
STF over all their space-time indices and transverse to the four
velocity, namely $u^\mu \,G_{\mu\mu_2\cdots\mu_\ell} = u^\mu
\,H_{\mu\mu_2\cdots\mu_\ell} = 0$, which is equivalent
to~\eqref{eq:G0i}.

Very important to the formalism is the fact that the Weyl tensor and its
covariant derivatives in~\eqref{eq:defGHcov} are to be evaluated at the
location of the particle $A$ following the regularization, as indicated by the
square brackets $[\cdots]_A$. Physically, the regularization is crucial
because it removes the self field of the particle $A$, and therefore permits
automatically selecting the external (tidal) field due to the other particles
$B\not= A$. We know one regularization able to give a complete, consistent and
physical answer in high PN approximations, namely dimensional regularization
(see \textit{e.g.}~\cite{DJSdim, BDE04}). In this paper, we shall systematically use
it. However, in our practical calculations at the relatively low NNL/2PN
order, it is simpler to use the Hadamard ``partie finie'' regularization,
since it has been shown~\cite{BiniDF12} to yield the same result for the
specific system we are interested in (see also discussions in
Ref.~\cite{BDE04}).

On the other hand, as argued in Refs.~\cite{BiniDF12, BDEI05dr}, we
can choose to use, for our purpose, the Riemann tensor instead of the
Weyl tensor in the definitions~\eqref{eq:defGHcov} of the tidal
moments. Indeed, the contributions due to the trace terms of the
Riemann tensor may be absorbed in the off-shell metric by redefining
it in a certain way. We give in~\cref{appendix:proof} a detailed proof
of this statement valid up to the NNL/2PN level.

Note finally that the tidal moments~\eqref{eq:defGHcov} have been
normalized in such a way that they admit a finite non-zero Newtonian
limit when $c\to+\infty$, and that the mass-type moments then match
those of Newtonian mechanics given in~\cref{appendix:Newtonian}. In
this limit, only the space components survive. We then get 
\begin{subequations}\label{eq:Nlimit}
\begin{align}
G_{L}^A &= \partial^A_{L} U_A +
\mathcal{O}\left(\frac{1}{c^{2}}\right)\,,\\ H_{L}^A &= 4
\,\varepsilon_{jk(i_\ell}\Bigl(\partial_{L-1)k}^A U^A_{j} + v_A^k
\partial^A_{L-1)j} U_A \Bigr) +
\mathcal{O}\left(\frac{1}{c^{2}}\right)\,,
\end{align}
\end{subequations}
where $\partial^A_{L}=\partial^A_{i_1}\cdots\partial^A_{i_\ell}$ with
$\partial^A_i=\partial/\partial y_A^i$; the potentials
$U_A=\sum_{B\not=A} G m_B/r_B$ and $U_A^i=\sum_{B\not=A} G m_B
v_B^i/r_B$ denote the Newtonian and gravitomagnetic potentials
regularized at the point $A$.

As the tidal moments are transverse to the velocity, the action~\eqref{eq:Sm}
can be rewritten in covariant form as
\begin{equation}\label{eq:Sm2}
S_{m} = \sum_{A} \int \dd \tau_{A} \left\{ - m_A c^2 + \sum_{\ell=2}^{+\infty}
  \frac{1}{2\ell!} \biggl[ \mu_{A}^{(\ell)}
  \, G^{A}_{\mu_1\cdots\mu_\ell} G_{A}^{\mu_1\cdots\mu_\ell} +
  \frac{\ell}{(\ell+1)c^{2}} \,\sigma_{A}^{(\ell)}
  \,H^{A}_{\mu_1\cdots\mu_\ell} H_{A}^{\mu_1\cdots\mu_\ell} \biggr]  + \cdots
\right\}\,.
\end{equation}
We observe that the reference to the local tetrad has completely
disappeared from the action. For convenience, we shall work only with
the global (tensorial) components $G_{\mu_1\cdots\mu_\ell}$ and
$H_{\mu_1\cdots\mu_\ell}$ of the moments henceforth.

The coefficients $\mu^{(\ell)}$ and $\sigma^{(\ell)}$ entering the
non-minimal action characterize the deformability and polarizability
of the body under the influence of the external tidal field. They are
linked to the dimensionless mass-type $k^{(\ell)}$ and current-type
$j^{(\ell)}$ second Love numbers as~\cite{BiniDF12}
\begin{equation}\label{eq:defpolarizability}
G \mu_{A}^{(\ell)} = \frac{2}{(2\ell-1)!!} \,k_{A}^{(\ell)}
R_{A}^{2\ell+1}\,,\qquad G \sigma_{A}^{(\ell)} =
\frac{\ell-1}{4(\ell+2)(2\ell-1)!!} \,j_{A}^{(\ell)} R_{A}^{2\ell+1}\,,
\end{equation}
where $R$ is the radius of the body (in a coordinate system such that the
area of the sphere of radius $R$ is $4\pi R^2$). In the effective description
\cref{eq:Sm2} of compact objects, only the coefficients $\mu_{A}^{(\ell)}$ and
$\sigma_{A}^{(\ell)}$ are measurable. The normalization constants in the first
equation~\eqref{eq:defpolarizability} are chosen to match usual Newtonian
definitions.

The polarizability coefficients~\eqref{eq:defpolarizability} actually
determine the formal PN order at which the tidal effects appear. For
compact objects, indeed, the compactness parameter defined as the
ratio $\mathcal{C}\sim G m/(R c^2)$ is of the order of one. Inserting
$\mathcal{C}\sim 1$ in~\cref{eq:defpolarizability}, we recover the
fact that the dominant tidal effect is due to the mass quadrupole and
is formally of the order of
\begin{equation}\label{eq:epstidal}
\epsilon_\text{tidal} \sim \frac{1}{c^{10}}\,,
\end{equation}
\textit{i.e.}, is comparable to a 5PN orbital effect. With the
notation~\eqref{eq:epstidal} for the dominant effect, we see that the
deformability coefficients in the action scale like
\begin{equation}\label{eq:order}
\left\{\mu_A^{(\ell)}\,,\,\sigma_A^{(\ell)}\right\} =
\mathcal{O}\left(\frac{\epsilon_\text{tidal}}{c^{4\ell-8}}\right)\,.
\end{equation}
As we aim at computing tidal effects up to the NNL/2PN order,
inspection of the action~\eqref{eq:Sm2} shows that we may consider
only the mass quadrupole, current quadrupole and mass octupole
interactions:
\begin{equation}\label{eq:Sm3}
S_{m} = \sum_{A} \int \dd \tau_{A} \left[ - m_A c^2 + \frac{\mu_{A}^{(2)}}{4}
  G^{A}_{\mu\nu}G_{A}^{\mu\nu} +
  \frac{\sigma_{A}^{(2)}}{6c^{2}}H^{A}_{\mu\nu}
   H_{A}^{\mu\nu} +
   \frac{\mu_{A}^{(3)}}{12} G^{A}_{\lambda\mu\nu}
   G_{A}^{\lambda\mu\nu}  +
   \mathcal{O}\left(\frac{\epsilon_\text{tidal}}{c^6}\right)\right]\,,
\end{equation}
where the specified remainder means that we neglect higher order --
NNNL and beyond -- terms. Direct application of the general scaling
relation~\eqref{eq:order} shows that
$\mu^{(2)}=\mathcal{O}(\epsilon_\text{tidal})$,
$\sigma^{(2)}=\mathcal{O}(\epsilon_\text{tidal})$, and
$\mu^{(3)}=\mathcal{O}(\epsilon_\text{tidal}/c^4)$. Thus, the first tidal
term in~\eqref{eq:Sm3} yields the leading effect together with NL and
NNL corrections, the second tidal term contains NL and NNL effects (because
of the explicit factor $1/c^2$ in the action), whereas the third one
represents a purely NNL effect.

\section{Metric and required elementary potentials}
\label{sec:metric}

To build an action for the sole matter variables, we (i) start from
the Einstein-Hilbert action~\eqref{eq:Sg} with the non-minimal matter
couplings~\eqref{eq:Sm}, (ii) solve the Einstein field equations
resulting from the metric variation by means of a direct PN iteration,
(iii) insert the explicit PN solution for the metric back into
Eqs.~\eqref{eq:Sg}--\eqref{eq:Sm}, which defines the so-called (PN)
Fokker action, say $S_\text{F}$. An important point is that, at the
NNL/2PN level, it is necessary and sufficient to insert into
Eqs.~\eqref{eq:Sg}--\eqref{eq:Sm} the metric generated by a system of
point particles, omitting all the terms associated with the body
internal structure.

To see this, we write, as in Ref.~\cite{BBBFMa}, the (allegedly
``exact'') PN solution of the Einstein field equations in terms of the
gothic metric deviation $h^{\mu\nu}=\sqrt{-g}g^{\mu\nu} -
\eta^{\mu\nu}$, using the particular vector variable
\begin{equation}\label{eq:varh}
  h = \bigl(h^{00ii}, h^{0i}, h^{ij}\bigr)\,,\quad\text{with}\quad
  h^{00ii} \equiv h^{00} + \delta_{ij}h^{ij}\,.
\end{equation}
We already know that the dominant tidal effect is due to the mass
quadrupole moment and pops up in the EoM at the
order~\eqref{eq:epstidal}. We can thus write the previous solution as
\begin{equation}\label{eq:hdecomp}
h = h_\text{pp} + h_\text{tidal}\,,
\end{equation}
where the first term is just the result for the metric generated by
point-particles (pp) without internal structure, and where the tidal
corrections therein are at least of the order of (with obvious
notation)
\begin{equation}\label{eq:htidal}
h_\text{tidal} = \mathcal{O}\left(\frac{\epsilon_\text{tidal}}{c^2},
  \frac{\epsilon_\text{tidal}}{c^3},
  \frac{\epsilon_\text{tidal}}{c^4}\right)\,.
\end{equation}
Since $h$ is an exact solution of the Einstein field equations we have
$\delta S_\text{F}/\delta h = 0$, which implies that the functional
derivative of the Fokker action evaluated for the ``approximate''
solution $h_\text{pp}$ will be of the order of the committed error,
namely (taking into account the coupling constant $c^4/(16\pi G)$ in
the field equations)
\begin{equation}\label{eq:dSdhpp}
\frac{\delta S_\text{F}}{\delta h} \bigl[h_\text{pp}\bigr] =
\mathcal{O}\left( c^2\,\epsilon_\text{tidal},
c\,\epsilon_\text{tidal}, \epsilon_\text{tidal}\right)\,.
\end{equation}
The two facts~\eqref{eq:htidal} and~\eqref{eq:dSdhpp} combined
together in a Taylor expansion of the action imply that
\begin{align}\label{eq:Spp}
S_\text{F} \bigl[h\bigr] &= S_\text{F} \bigl[h_\text{pp}\bigr] + \int
\dd^4x\,\frac{\delta S_\text{F}}{\delta h} \bigl[h_\text{pp}\bigr]
\,h_\text{tidal} + \mathcal{O}\left( h^2_\text{tidal}\right)
\nonumber\\ &= S_\text{F} \bigl[h_\text{pp}\bigr] + \mathcal{O}\left(
\epsilon_\text{tidal}^2\right)\,,
\end{align}
and we conclude that the final remainder $\mathcal{O}(
\epsilon_\text{tidal}^2)$ is at least comparable to a 10PN effect
$\mathcal{O}( c^{-20})$ [see~\cref{eq:epstidal}]. Therefore, it is amply
sufficient to insert into the Fokker action the metric  $h_\text{pp}$ for point
particles without internal structure. We can
recover this conclusion from a general statement proved in
Ref.~\cite{BBBFMa} and called the method ``$n+2$'', according to
which, in order to control the Fokker action at some $n$PN order, it
is necessary and sufficient to insert the components of the metric $h$
with all the PN corrections up to the order $1/c^{n+2}$ included. In
our case, we want the Fokker action up to the NNL order, which means
formally 7PN, hence $n=7$, so that we require the metric up to the maximal
order $1/c^{9}$ while tidal effects are of higher order
[see~\cref{eq:htidal}]. The same argument has also been shown and used
in Sec.~II.E of Ref.~\cite{BiniDF12}.

In this paper, we shall not try to compute the full action, including all the
terms up to NNL order $\mathcal{O}(\epsilon_\text{tidal}/c^4)$, but only the
tidal NNL contributions therein, proceeding essentially as in
Ref.~\cite{BiniDF12} although staying in harmonic coordinates. Consequently, we shall
need the point-particle metric up to the 2PN order only, so as to obtain the
regularized Weyl or Riemann tensor of point particles at the 2PN order, which
is the minimum requirement to control the tidal moments at the same accuracy
level:
\begin{subequations}\label{eq:tidalR}
\begin{align}
  G^A_{\mu\nu} &= - c^2 \bigl[R_{\mu\rho\nu\sigma}\bigr]_A
  u_A^{\rho}u_A^{\sigma}\,, \\ H^A_{\mu\nu} &= 2 c^3 \bigl[
    R^{*}_{(\mu\underline{\rho}\nu)\sigma}\bigr]_A
  u_A^{\rho}u_A^{\sigma}\,, \\ G^A_{\lambda\mu\nu} &= - c^2 \bigl[
    \nabla^\perp_{(\lambda}
    \,R_{\mu\underline{\rho}\nu)\sigma}\bigr]_A
  u_A^{\rho}u_A^{\sigma}\,.
\end{align}
\end{subequations}
Remind that, for this calculation, the Weyl and the Riemann tensors give an
equivalent dynamics (see~\cref{appendix:proof}). On the other hand, one can
show that replacing the STF operator by the symmetrization operator in the
definitions~\eqref{eq:defGHcov} for the mass quadrupole, current quadrupole
and mass octupole moments does not affect the values of those tensors. The
resulting expressions, provided in~\cref{appendix:proof}, are simpler than the
original formulae. The tensors~\eqref{eq:tidalR} are then obtained by
substituting there the Riemann tensor to the Weyl one. However, the off-shell
mass-type tidal moments defined in this manner are no longer trace-free,
contrary to their Weyl counterparts.

At the 2PN order, the metric of a general matter system in harmonic
coordinates can be parametrized by the set of potentials
$\{V,V_{i},\hat{W}_{ij},\hat{R}_{i},\hat{X}\}$ in the following way:
\begin{subequations}\label{eq:metric}
\begin{align}
g_{00} &= -1 + \frac{2V}{c^{2}} - \frac{2V^{2}}{c^{4}}+ \frac{8}{c^{6}}
\left(\hat{X} + V_{i}V_{i} + \frac{V^{3}}{6} \right) +
\mathcal{O}\left( \frac{1}{c^{8}} \right)\,, \\ g_{0i} &=
-\frac{4V_{i}}{c^{3}} - \frac{8\hat{R}_{i}}{c^{5}} + \mathcal{O}\left(
\frac{1}{c^{7}} \right)\,, \\ g_{ij} &= \delta_{ij}\left(1 +
\frac{2V}{c^{2}} + \frac{2V^{2}}{c^{4}} \right) +
\frac{4\hat{W}_{ij}}{c^{4}} + \mathcal{O}\left( \frac{1}{c^{6}}
\right)\,.
\end{align}
\end{subequations}
These potentials admit a non-zero finite Newtonian limit and solve the
flat-space wave equations (with
$\Box=\eta^{\mu\nu}\partial^2_{\mu\nu}$)
\begin{subequations}\label{eq:potential}
\begin{align}
\Box V &= -4 \pi G \sigma\,, \\ 
\Box V_{i} &= -4  \pi G \sigma_{i}\,, \\ 
\Box \hat{W}_{ij} &= -4\pi G\left(\sigma_{ij} - \delta_{ij}
   \sigma\indices{_k_k} \right) -\partial_{i}V
   \partial_{j}V\,, \\
\Box \hat{R}_{i} &= -4 \pi G \left(V \sigma_{i} - V_{i} \sigma \right) -2
  \partial_{k}V \partial_{i}V_{k} - \frac{3}{2}\partial_{t}V
  \partial_{i}V\,, \\
\Box \hat{X} &= -4 \pi G V \sigma\indices{_k_k} +
   2V_{k}\partial_{t}\partial_{k}V + V \partial_{t}^{2}V
  +\frac{3}{2}(\partial_{t}V)^{2} -
  2\partial_{i}V_{j}\partial_{j}V_{i}
  +\hat{W}_{ij}\partial_{ij}V\,,
\end{align}
\end{subequations}
where the matter source densities are defined in terms of the
components of the matter stress-energy tensor as
\begin{equation}\label{eq:source}
\sigma = \frac{T^{00}+T^{ii}}{c^2}\,,\qquad \sigma_{i} =
\frac{T^{0i}}{c}\,,\qquad \sigma_{ij} = T^{ij}\,,
\end{equation}
with $T^{ii}=\delta_{ij}T^{ij}$. To perform a consistent Fokker
reduction of the original action, the solutions of
Eqs.~\eqref{eq:potential} must be in principle constructed with the
symmetric Green function, which kills all contributions of odd powers
of $1/c$ at the current approximation level. As discussed above,
thanks to the properties of the Fokker action, we only need the
metric produced by point-like particles and can neglect tidal effects
when inserting the metric~\eqref{eq:metric} into the Fokker action.
Therefore, we shall just compute the potentials for point particles
without including any internal structure effect. The requested
potentials have already been published elsewhere~\cite{BFP98}, except
that we compute here their off-shell values, without replacement of
accelerations by means of the EoM (we then call them the ``unreduced''
potentials). However, it is known that the replacement of
accelerations in the action is equivalent to performing an unphysical
shift of the particles' worldlines~\cite{S84}. We have checked that, indeed, by
inserting the reduced (``on-shell'') versions of the potentials into
the action, the final gauge invariant result for the conserved energy
reduced to circular orbits, which we shall obtain below [in~\cref{eq:Ex}],
comes out the same.

For point particles without spins the matter source
terms~\eqref{eq:source} take the form
\begin{subequations}\label{eq:potentiel}
\begin{align}
\sigma(\mathbf{x},t) &= \sum_{A} \tilde{\mu}_{A}(t)\,\delta^{(3)}
\bigl(\mathbf{x}-\bm{y}_{A}(t)\bigr)\,,\\ \sigma_i(\mathbf{x},t) &=
\sum_{A} \mu_{A}\,v_A^i\,\delta^{(3)}
\bigl(\mathbf{x}-\bm{y}_{A}(t)\bigr)\,,\\ \sigma_{ij}(\mathbf{x},t) &=
\sum_{A} \mu_{A}\,v_A^i v_A^j\,\delta^{(3)}
\bigl(\mathbf{x}-\bm{y}_{A}(t)\bigr)\,,
\end{align}
\end{subequations}
where the three-dimensional Dirac function is confined to the worldline
$\bm{y}_A(t)$ and we pose for the effective time-varying masses (with $m_A$
the constant PN mass)
\begin{equation}\label{eq:mu}
\mu_A(t) = \frac{m_A\,c}{\sqrt{[g\,g_{\mu\nu}]_A v_A^\mu v_A^\nu}} \,,\qquad
\tilde{\mu}_A = \left(1 + \frac{\bm{v}_A^2}{c^2}\right)\mu_A\,,
\end{equation}
In Eqs.~\eqref{eq:potentiel}--\eqref{eq:mu}, the worldlines are
parametrized by the coordinate time $t=x^0/c$ of the harmonic
coordinates; the coordinate velocities are $v_{A}^{\mu}=(c,v_A^i)$,
with $v_A^i=c u_A^i/u_A^0 = \dd y_A^i/\dd t$, and the relativistic
Lorentz factor reads $u_{A}^{0}=(-[g_{\mu \nu}]_A
v_{A}^{\mu}v_{A}^{\nu}/c^{2})^{-1/2}$. The metric is computed at the
location of the particle $A$ following dimensional regularization; in
particular, we have $[g\,g_{\mu\nu}]_A = [g]_A\,[g_{\mu\nu}]_A$
in~\eqref{eq:mu}. As we said, in practical calculations, we use the
Hadamard regularization, which is equivalent to dimensional
regularization up to the relatively low NNL/2PN order~\cite{BDE04,BiniDF12}.\\

To summarize, the metric in \cref{eq:metric} is not the full metric and does
not contain tidal effects. However, as proved in
\cref{eq:varh,eq:hdecomp,eq:htidal,eq:dSdhpp,eq:Spp}, it is sufficient to
insert into the Fokker action $S_g + S_m$ needed to obtain the tidal effects
in the equations of motion at the requested order. In our work, the full
metric is not controlled. Indeed, we only compute the equations of motion of
the particles (and the associated conserved quantities). With the Fokker
method, we do not need nor have access to the full metric outside the
particle's worldlines.

\section{Tidal effects in the equations of motion to NNL
  order} \label{sec:Fokker}

From the discussion in the previous section, we know that, up to the
NNL order, the only terms in the Fokker action that depend on the
bodies' internal structure are those that are explicitly present into
the matter action~\eqref{eq:Sm3}. Here, we provide the results for the
(coordinate basis components of the) tidal mass-quadrupole,
mass-octupole and current-quadrupole moments at the NNL order felt by
the body 1, \textit{i.e.}, regularized at the point 1. We
find\footnote{The notation $r_{12}=\vert \bm{y}_{1}-\bm{y}_{2}\vert$
  represents the Euclidean distance between the two bodies (at
  constant time $y_{1}^{0}=y_{2}^{0}=c\, t$); the unit direction from
  body 2 to body 1 is then $n_{12}^{i} =
  (y_{1}^{i}-y_{2}^{i})/r_{12}$; $v_{12}^{i}=v_{1}^i-v_{2}^i$ stands
  for the relative velocity; the usual Euclidean scalar product of
  vectors is denoted with parentheses, \textit{e.g.}
  $(n_{12}v_1)=\bm{n}_{12}\cdot\bm{ v}_1$; the cross product is
  denoted, \textit{e.g.}  $(n_{12}\times v_{12})_i$, and the mixed
  product, \textit{e.g.} $(n_{12}, v_1, v_2)= (n_{12} \ v_1 \times
  v_2)$. All calculations are done with the software Mathematica and
  the tensor package \textit{xAct}~\cite{xtensor}.}
\begin{subequations}\label{eq:tidalmoments}
\begin{align}
[G_{ij}]_1 &= \frac{G m_{2}}{r_{12}^3} \Biggl[3 n_{12\langle i} n_{12j\rangle}
 + \frac{1}{c^{2}} \biggl [n_{12\langle i} n_{12j\rangle} \Bigl(- \frac{15}{2}
             (n_{12}{} v_{2}{})^2
 + 6 v_{12}{}^{2}
 -  \frac{3}{2} r_{12} (n_{12}{} a_{2}{})
 -  \frac{3 G m_{1}}{r_{12}}
 -  \frac{3 G m_{2}}{r_{12}}\Bigr)\nonumber\\
& - 6 n_{12\langle i} v_{1j\rangle} (n_{12}{} v_{12}{})
 + 2 v_{1\langle i} v_{1j\rangle}
 + n_{12\langle i} v_{2j\rangle} \Bigl(12 (n_{12}{} v_{1}{})
 - 6 (n_{12}{} v_{2}{})\Bigr)
 - 6 v_{1\langle i} v_{2j\rangle}
 + 3 v_{2\langle i} v_{2j\rangle}
 -  3 a_{2\langle i} n_{12j\rangle} r_{12}\nonumber\\
& + \delta_{ij} \Bigl((n_{12}{} v_{1}{})^2
 -  \frac{1}{3} v_{1}{}^{2}\Bigr)\biggl]
 + \frac{1}{c^{4}} \biggl\{n_{12\langle i} n_{12j\rangle} \biggl
[\frac{105}{8} (n_{12}{} v_{2}{})^4
 + 30 (n_{12}{} v_{2}{})^2 (v_{1}{} v_{2}{})
 + 6 (v_{1}{} v_{2}{})^2
 - 15 (n_{12}{} v_{2}{})^2 v_{1}{}^{2}\nonumber\\
& - 12 (v_{1}{} v_{2}{}) v_{1}{}^{2}
 + 6 v_{1}{}^{4}
 -  \frac{45}{2} (n_{12}{} v_{2}{})^2 v_{2}{}^{2}
 - 12 (v_{1}{} v_{2}{}) v_{2}{}^{2}
 + 6 v_{1}{}^{2} v_{2}{}^{2}
 + 6 v_{2}{}^{4}
 + G m_{2} (n_{12}{} a_{2}{})\nonumber\\
& + \frac{G m_{1}}{r_{12}} \Bigl(- \frac{291}{2} (n_{12}{} v_{1}{})^2
 + 291 (n_{12}{} v_{1}{}) (n_{12}{} v_{2}{})
 -  \frac{273}{2} (n_{12}{} v_{2}{})^2
 + 35 v_{12}{}^{2}\Bigr)
 + G m_{1} \Bigl(14 (n_{12}{} a_{1}{})
 - 10 (n_{12}{} a_{2}{})\Bigr)\nonumber\\
& + \frac{G m_{2}}{r_{12}} \Bigl(9 (n_{12}{} v_{2}{})^2
 + 18 v_{12}{}^{2}\Bigr)
 + \frac{1}{8} r_{12}^3 (\ddot{a}_{2} n_{12}{})
 -  \frac{15 G^2 m_{1}^2}{14 r_{12}^2}
 + \frac{35 G^2 m_{1} m_{2}}{r_{12}^2}
 + \frac{5 G^2 m_{2}^2}{r_{12}^2}
 + r_{12} \Bigl(12 (v_{1}{} a_{2}{}) (n_{12}{} v_{2}{})\nonumber\\
& -  \frac{27}{2} (v_{2}{} a_{2}{}) (n_{12}{} v_{2}{})
 + \frac{45}{4} (n_{12}{} a_{2}{}) (n_{12}{} v_{2}{})^2
 + 6 (n_{12}{} a_{2}{}) (v_{1}{} v_{2}{})
 - 3 (n_{12}{} a_{2}{}) v_{1}{}^{2}
 -  \frac{9}{2} (n_{12}{} a_{2}{}) v_{2}{}^{2}\Bigr)\nonumber\\
& + r_{12}^2 \Bigl(\frac{9}{8} (n_{12}{} a_{2}{})^2
 -  \frac{15}{8} a_{2}{}^{2}
 + \frac{3}{2} (n_{12}{} v_{2}{}) (n_{12}{} \dot{a}_{2})
 + 2 (v_{1}{} \dot{a}_{2})
 - 2 (v_{2}{} \dot{a}_{2})\Bigr)\biggl]
 + n_{12\langle i} v_{1j\rangle} \biggl [\frac{62 G m_{1}}{r_{12}} (n_{12}{}
v_{12}{})\nonumber\\
& -  \frac{18 G m_{2}}{r_{12}} (n_{12}{} v_{12}{})
 + 15 (n_{12}{} v_{1}{}) (n_{12}{} v_{2}{})^2
 - 15 (n_{12}{} v_{2}{})^3
 + 6 (n_{12}{} v_{2}{}) (v_{1}{} v_{2}{})
 + 6 (n_{12}{} v_{2}{}) v_{12}{}^{2}
 - 6 (n_{12}{} v_{1}{}) v_{1}{}^{2}\nonumber\\
& + r_{12} \Bigl(- (v_{12}{} a_{2}{})
 + 3 (n_{12}{} a_{2}{}) (n_{12}{} v_{1}{})
 - 9 (n_{12}{} a_{2}{}) (n_{12}{} v_{2}{})\Bigr)
 -  r_{12}^2 (n_{12}{} \dot{a}_{2})\biggl]
 + v_{1\langle i} v_{1j\rangle} \Bigl(-3 (n_{12}{} v_{2}{})^2
 + 2 v_{1}{}^{2}\nonumber\\
& -  r_{12} (n_{12}{} a_{2}{})
 -  \frac{3 G m_{1}}{r_{12}}
 + \frac{6 G m_{2}}{r_{12}}\Bigr)
 + n_{12\langle i} v_{2j\rangle} \biggl [-30 (n_{12}{} v_{1}{}) (n_{12}{} v_{2}{})^2
 + 15 (n_{12}{} v_{2}{})^3
 - 12 (n_{12}{} v_{1}{}) (v_{1}{} v_{2}{})\nonumber\\
& + 12 (n_{12}{} v_{1}{}) v_{1}{}^{2}
 + 12 (n_{12}{} v_{1}{}) v_{2}{}^{2}
 - 6 (n_{12}{} v_{2}{}) v_{2}{}^{2}
 + \frac{G m_{1}}{r_{12}} \Bigl(-68 (n_{12}{} v_{1}{})
 + 62 (n_{12}{} v_{2}{})\Bigr)
 + \frac{G m_{2}}{r_{12}} \Bigl(12 (n_{12}{} v_{1}{})\nonumber\\
& - 18 (n_{12}{} v_{2}{})\Bigr)
 + r_{12}^2 (n_{12}{} \dot{a}_{2})
 + r_{12} \Bigl(-2 (v_{1}{} a_{2}{})
 - 6 (n_{12}{} a_{2}{}) (n_{12}{} v_{1}{})
 -  (v_{2}{} a_{2}{})
 + 9 (n_{12}{} a_{2}{}) (n_{12}{} v_{2}{})\Bigr)\biggl]\nonumber\\
& + v_{1\langle i} v_{2j\rangle} \Bigl(-6 (n_{12}{} v_{1}{}) (n_{12}{} v_{2}{})
 + 15 (n_{12}{} v_{2}{})^2
 - 6 (v_{1}{} v_{2}{})
 - 6 v_{12}{}^{2}
 + 5 r_{12} (n_{12}{} a_{2}{})
 + \frac{8 G m_{1}}{r_{12}}
 -  \frac{10 G m_{2}}{r_{12}}\Bigr)\nonumber\\
& + v_{2\langle i} v_{2j\rangle} \Bigl(6 (n_{12}{} v_{1}{})^2
 -  \frac{15}{2} (n_{12}{} v_{2}{})^2
 + 3 v_{2}{}^{2}
 -  \frac{5}{2} r_{12} (n_{12}{} a_{2}{})
 -  \frac{4 G m_{1}}{r_{12}}
 + \frac{5 G m_{2}}{r_{12}}\Bigr)
 + 4 G m_{1} a_{1\langle i} n_{12j\rangle}\nonumber\\
   & + a_{2\langle i} n_{12j\rangle} \biggl [r_{12} \Bigl(-12 (n_{12}{} v_{1}{})
   (n_{12}{} v_{2}{})
 + \frac{27}{2} (n_{12}{} v_{2}{})^2
 + 4 (v_{1}{} v_{2}{})
 - 2 v_{1}{}^{2}
 - 5 v_{2}{}^{2}\Bigr)
 + \frac{9}{2} r_{12}^2 (n_{12}{} a_{2}{})
 -  3 G m_{1}\nonumber\\
& -  G m_{2}\biggl]
 + a_{2\langle i} v_{1j\rangle} r_{12} \Bigl(- (n_{12}{} v_{1}{})
 + 7 (n_{12}{} v_{2}{})\Bigr)
 + a_{2\langle i} v_{2j\rangle} r_{12} \Bigl(-2 (n_{12}{} v_{1}{})
 - 7 (n_{12}{} v_{2}{})\Bigr)
 -  \frac{5}{4} a_{2\langle i} a_{2j\rangle} r_{12}^2\nonumber\\
& + n_{12\langle i} \dot{a}_{2j\rangle} r_{12}^2 \Bigl(-2 (n_{12}{} v_{1}{})
 + 5 (n_{12}{} v_{2}{})\Bigr)
 + 3 v_{1\langle i} \dot{a}_{2j\rangle} r_{12}^2
 -  3 v_{2\langle i} \dot{a}_{2j\rangle} r_{12}^2
 + \frac{7}{4} n_{12\langle i} \ddot{a}_{2j\rangle} r_{12}^3
+ \delta_{ij} \biggl [- \frac{5}{2} (n_{12}{} v_{1}{})^2
(n_{12}{} v_{2}{})^2\nonumber\\
& - 2 (n_{12}{} v_{1}{}) (n_{12}{} v_{2}{}) (v_{1}{} v_{2}{})
 + (v_{1}{} v_{2}{})^2
 + (n_{12}{} v_{1}{})^2 v_{1}{}^{2}
 + \frac{3}{2} (n_{12}{} v_{2}{})^2 v_{1}{}^{2}
 -  \frac{1}{3} v_{1}{}^{4}
 + 2 (n_{12}{} v_{1}{})^2 v_{2}{}^{2}
 -  v_{1}{}^{2} v_{2}{}^{2}\nonumber\\
& -  \frac{4}{3} G m_{1} (n_{12}{} a_{2}{})
 + \frac{G m_{1}}{r_{12}} \Bigl(- \frac{16}{3} (n_{12}{} v_{12}{})^2
 -  (n_{12}{} v_{1}{})^2
 + \frac{4}{3} v_{12}{}^{2}
 + \frac{1}{3} v_{1}{}^{2}\Bigr)
 + \frac{G m_{2}}{r_{12}} \Bigl(4 (n_{12}{} v_{12}{})^2
 -  (n_{12}{} v_{1}{})^2\nonumber\\
& -  \frac{4}{3} v_{12}{}^{2}
 + \frac{1}{3} v_{1}{}^{2}\Bigr)
 + r_{12} \Bigl(\frac{4}{3} (v_{12}{} a_{2}{}) (n_{12}{} v_{12}{})
 -  (v_{1}{} a_{2}{}) (n_{12}{} v_{1}{})
 -  \frac{1}{2} (n_{12}{} a_{2}{}) (n_{12}{} v_{1}{})^2
 + \frac{1}{2} (n_{12}{} a_{2}{}) v_{1}{}^{2}\Bigr)\nonumber\\
& -  \frac{16 G^2 m_{1} m_{2}}{3 r_{12}^2}
 + \frac{2 G^2 m_{2}^2}{3 r_{12}^2}
 + r_{12}^2 \Bigl(\frac{4}{3} a_{2}{}^{2}
 -  \frac{4}{3} (v_{1}{} \dot{a}_{2})
 + \frac{4}{3} (v_{2}{} \dot{a}_{2})\Bigr)\biggl]\biggl\}\Biggl]+
\mathcal{O}\left(\frac{1}{c^6}\right)\, ,\\
[H_{ij}]_1 &=\frac{G m_{2}}{r_{12}^3} \biggl\{12 (n_{12}{}\times
             v_{12}{})_{\langle i}n_{12j\rangle}
 + \frac{1}{c^{2}} \biggl [(n_{12}{}\times v_{12}{})_{\langle i}n_{12j\rangle}
             \Bigl(-30 (n_{12}{} v_{2}{})^2
 + 12 (v_{1}{} v_{2}{})
 + 12 v_{12}{}^{2}
 -  6 r_{12} (n_{12}{} a_{2}{})\nonumber\\
& + \frac{4 G m_{1}}{r_{12}}
 + \frac{12 G m_{2}}{r_{12}}\Bigr)
 -  12 (a_{2}{}\times n_{12}{})_{\langle i}n_{12j\rangle} r_{12} (n_{12}{} v_{2}{})
 + 12 (n_{12}{}\times v_{12}{})_{\langle i}v_{2j\rangle} (n_{12}{} v_{1}{})
 -  2 (a_{2}{}\times v_{12}{})_{\langle i}n_{12j\rangle} r_{12}\nonumber\\
& -  2 a_{2\langle i}(n_{12}{}\times v_{12}{})_{j\rangle} r_{12}
 + 2 (n_{12}{}\times \dot{a}_{2}{})_{\langle i}n_{12j\rangle} r_{12}^2
 + 4 \delta_{ij} (n_{12}{},v_{1}{},v_{2}{}) (n_{12}{}
 v_{1}{})\biggl]\biggl\} +
\mathcal{O}\left(\frac{1}{c^4}\right)\, , \\
[G_{ijk}]_1&=- \frac{15 G m_{2} n_{12\langle i} n_{12j}
         n_{12k\rangle}}{r_{12}^4}+ \mathcal{O}\left( \frac{1}{c^2}\right)\,.
\end{align}
\end{subequations}
The other components of the tidal moments are readily obtained from,
\textit{e.g.}, the relations $[G_{0i}]_1 = -v_1^j \,[G_{ij}]_1/c$ and
$[G_{00}]_1 = v_1^i v_1^j \,[G_{ij}]_1/c^2$, which are equivalent to
$[G_{\tet{0}\tet{0}}]_1 = [G_{\tet{0}\tet{i}}]_1 = 0$ in tetradic notation. In
Eqs.~\eqref{eq:tidalmoments}, most of the terms are STF, which we denote by
angular brackets surounding the indices. Note however, as mentioned
in~\cref{sec:metric}, the appearance of pure trace contributions, due to the
fact that we have not resorted here to tetradic projections and have used the
Riemann tensor instead of the Weyl tensor [see the discussion
in~\cref{appendix:proof}].

With the latter results and the 2PN metric~\eqref{eq:metric} in hands, it is
straightforward to get the Lagrangian up to the relative NNL/2PN
order for the finite-size tidal contributions. As usual, we apply a
number of procedures to eliminate multiple time derivatives of the
accelerations and reduce the numbers of terms, in particular removing
those that contain higher time derivatives of the
accelerations by adding suitable double-zero terms and total time
derivatives~\cite{DS85}. Recalling our notation introduced
in~\cref{eq:hdecomp}, we write
\begin{equation}\label{eq:Lpptidal}
L = L_\text{pp} + L_\text{tidal}\,,
\end{equation}
where, to be consistent with the NNL order truncation, we recall here the
Lagrangian for point particles up to 2PN order in harmonic
coordinates, which is a generalized Lagrangian depending on positions
$y_A^i(t)$, velocities $v_A^i(t)$, as well as accelerations
$a_A^i(t)=\dd v_A^i/\dd t$ (see, \textit{e.g.}, Eq.~(209)
of~\cite{BlanchetLR}):
\begin{align}\label{eq:Lpp}
L_\text{pp} &= \frac{m_1 v_1^2}{2} + \frac{G m_1 m_2}{2 r_{12}}
  \nonumber \\ & 
+ \frac{1}{c^2} \left\{ - \frac{G^2 m_1^2 m_2}{2
    r_{12}^2} + \frac{m_1 v_1^4}{8} + \frac{G m_1 m_2}{r_{12}} \left(
   - \frac{1}{4} (n_{12}v_1) (n_{12}v_2) + \frac{3}{2} v_1^2 -
  \frac{7}{4} (v_1v_2) \right)  \right\} \nonumber \\ & 
+ \frac{1}{c^4} \Bigg\{ \frac{G^3 m_1^3 m_2}{2 r_{12}^3} + \frac{19
  G^3 m_1^2 m_2^2}{8 r_{12}^3} \nonumber \\ & \qquad ~\, + \frac{G^2
  m_1^2 m_2}{r_{12}^2} \left( \frac{7}{2} (n_{12}v_1)^2 - \frac{7}{2}
(n_{12}v_1) (n_{12}v_2) + \frac{1}{2}(n_{12}v_2)^2 + \frac{1}{4} v_1^2
- \frac{7}{4} (v_1v_2) + \frac{7}{4} v_2^2 \right) \nonumber \\ &
\qquad ~\, + \frac{G m_1 m_2}{r_{12}} \bigg( \frac{3}{16}
(n_{12}v_1)^2 (n_{12}v_2)^2 - \frac{7}{8} (n_{12}v_2)^2 v_1^2 +
\frac{7}{8} v_1^4 + \frac{3}{4} (n_{12}v_1) (n_{12}v_2) (v_1v_2)
\nonumber \\ & \qquad \qquad \qquad \qquad - 2 v_1^2 (v_1v_2) +
\frac{1}{8} (v_1v_2)^2 + \frac{15}{16} v_1^2 v_2^2 \bigg) + \frac{m_1
  v_1^6}{16} \nonumber \\ & \qquad ~\, + G m_1 m_2 \left( -
\frac{7}{4} (a_1 v_2) (n_{12}v_2) - \frac{1}{8} (n_{12} a_1)
(n_{12}v_2)^2 + \frac{7}{8} (n_{12} a_1) v_2^2 \right) \Bigg\} +
1\leftrightarrow 2 + \mathcal{O}\left(\frac{1}{c^5}\right)\,.
\end{align}
To the terms given above, we must add their symmetric counterpart in
the exchange of the two particles, as indicated by the notation $1\leftrightarrow
2$. Now, the main result of the present paper is the complete
expression of the tidal part of the Lagrangian up to the NNL/2PN order
in harmonic coordinates. It reads
\begin{align}\label{eq:Ltidal}
L_{\text{tidal}}&= \frac{G^2 m_{2}^2}{r_{12}^6} \biggl\{\frac{3}{2} \mu_1^{(2)}
 + \frac{1}{c^{2}} \biggl [\mu_1^{(2)} \Bigl(- \frac{9}{2} (n_{12}{} v_{1}{})^2
 - 18 (n_{12}{} v_{1}{}) (n_{12}{} v_{2}{})
 + 18 (n_{12}{} v_{2}{})^2
 -  \frac{9}{2} (v_{1}{} v_{2}{})
 + \frac{15}{4} v_{1}{}^{2}\Bigr)\nonumber\\
& + \sigma_1^{(2)} \Bigl(-12 (n_{12}{} v_{12}{})^2
 + 12 v_{12}{}^{2}\Bigr)
 -  \frac{3 G m_{1} \mu_1^{(2)}}{r_{12}}
 -  \frac{21 G m_{2} \mu_1^{(2)}}{2 r_{12}}\biggl]
 + \frac{1}{c^{4}} \biggl [\mu_1^{(2)}
   \Bigl(\frac{9}{2} (n_{12}{} v_{1}{})^4\nonumber\\
& - 18 (n_{12}{} v_{1}{})^3 (n_{12}{} v_{2}{})
 + 45 (n_{12}{} v_{1}{})^2 (n_{12}{} v_{2}{})^2
 - 54 (n_{12}{} v_{1}{}) (n_{12}{} v_{2}{})^3
 + \frac{63}{2} (n_{12}{} v_{2}{})^4
 + 9 (n_{12}{} v_{1}{}) (n_{12}{} v_{2}{}) (v_{1}{} v_{2}{})\nonumber\\
& - 18 (n_{12}{} v_{2}{})^2 (v_{1}{} v_{2}{})
 + \frac{9}{2} (v_{1}{} v_{2}{})^2
 - 9 (n_{12}{} v_{1}{})^2 v_{12}{}^{2}
 + 27 (n_{12}{} v_{1}{}) (n_{12}{} v_{2}{}) v_{12}{}^{2}
 - 36 (n_{12}{} v_{2}{})^2 v_{12}{}^{2}\nonumber\\
& + 9 (v_{1}{} v_{2}{}) v_{12}{}^{2}
 + 9 v_{12}{}^{4}
 -  \frac{9}{4} (n_{12}{} v_{1}{})^2 v_{1}{}^{2}
 -  \frac{9}{2} (n_{12}{} v_{1}{}) (n_{12}{} v_{2}{}) v_{1}{}^{2}
 + \frac{27}{2} (n_{12}{} v_{2}{})^2 v_{1}{}^{2}
 - 9 (v_{1}{} v_{2}{}) v_{1}{}^{2}\nonumber\\
& -  \frac{27}{4} v_{12}{}^{2} v_{1}{}^{2}
 + \frac{69}{16} v_{1}{}^{4}\Bigr)
 + \mu_1^{(2)} r_{12} \Bigl(-12 (v_{12}{} a_{2}{}) (n_{12}{} v_{1}{})
 + 60 (n_{12}{} a_{2}{}) (n_{12}{} v_{1}{})^2
 + 21 (v_{12}{} a_{2}{}) (n_{12}{} v_{2}{})\nonumber\\
& -  \frac{9}{2} (v_{1}{} a_{2}{}) (n_{12}{} v_{2}{})
 - 102 (n_{12}{} a_{2}{}) (n_{12}{} v_{1}{}) (n_{12}{} v_{2}{})
 + 60 (n_{12}{} a_{2}{}) (n_{12}{} v_{2}{})^2
 + \frac{69}{2} (n_{12}{} a_{2}{}) (v_{1}{} v_{2}{})
 -  \frac{69}{4} (n_{12}{} a_{2}{}) v_{1}{}^{2}\nonumber\\
& -  \frac{39}{2} (n_{12}{} a_{2}{}) v_{2}{}^{2}\Bigr)
 + \sigma_1^{(2)} \Bigl(60 (n_{12}{} v_{12}{})^4
 - 96 (n_{12}{} v_{12}{})^3 (n_{12}{} v_{1}{})
 + 48 (n_{12}{} v_{12}{})^2 (n_{12}{} v_{1}{})^2
 - 24 (n_{12}{} v_{12}{})^2 (v_{1}{} v_{2}{})\nonumber\\
& + 24 (n_{12}{} v_{12}{}) (n_{12}{} v_{1}{}) (v_{1}{} v_{2}{})
 + 12 (v_{1}{} v_{2}{})^2
 - 84 (n_{12}{} v_{12}{})^2 v_{12}{}^{2}
 + 96 (n_{12}{} v_{12}{}) (n_{12}{} v_{1}{}) v_{12}{}^{2}
 - 36 (n_{12}{} v_{1}{})^2 v_{12}{}^{2}\nonumber\\
& + 24 (v_{1}{} v_{2}{}) v_{12}{}^{2}
 + 24 v_{12}{}^{4}
 + 18 (n_{12}{} v_{12}{})^2 v_{1}{}^{2}
 - 24 (n_{12}{} v_{12}{}) (n_{12}{} v_{1}{}) v_{1}{}^{2}
 - 24 (v_{1}{} v_{2}{}) v_{1}{}^{2}
 - 18 v_{12}{}^{2} v_{1}{}^{2}
 + 12 v_{1}{}^{4}\Bigr)\nonumber\\
& + \sigma_1^{(2)} r_{12} \Bigl(16 (n_{12}{} a_{2}{}) (n_{12}{} v_{12}{})^2
 + 24 (v_{12}{} a_{2}{}) (n_{12}{} v_{1}{})
 - 24 (n_{12}{} a_{2}{}) (n_{12}{} v_{12}{}) (n_{12}{} v_{1}{})
 - 16 (n_{12}{} a_{2}{}) v_{12}{}^{2}\Bigr)\nonumber\\
& + \frac{G m_{1} \mu_1^{(2)}}{r_{12}} \Bigl(\frac{807}{8} (n_{12}{} v_{1}{})^2
 + \frac{381}{8} (n_{12}{} v_{1}{}) (n_{12}{} v_{2}{})
 - 138 (n_{12}{} v_{2}{})^2
 -  \frac{387}{8} (v_{1}{} v_{2}{})
 + \frac{63}{8} v_{1}{}^{2}
 + 42 v_{2}{}^{2}\Bigr)\nonumber\\
& + \frac{G m_{2} \mu_1^{(2)}}{r_{12}} \Bigl(\frac{27}{2} (n_{12}{} v_{1}{})^2
 + \frac{1051}{8} (n_{12}{} v_{1}{}) (n_{12}{} v_{2}{})
 -  \frac{865}{8} (n_{12}{} v_{2}{})^2
 + \frac{83}{8} (v_{1}{} v_{2}{})
 -  \frac{45}{4} v_{1}{}^{2}
 + \frac{49}{8} v_{2}{}^{2}\Bigr)\nonumber\\
& + \frac{G m_{1} \sigma_1^{(2)}}{r_{12}} \Bigl(-8 (n_{12}{} v_{12}{})^2
 + 8 v_{12}{}^{2}\Bigr)
 + \frac{G m_{2} \sigma_1^{(2)}}{r_{12}} \Bigl(36 (n_{12}{} v_{12}{})^2
 - 36 v_{12}{}^{2}\Bigr)
 -  \frac{60 G^2 m_{1}^2 \mu_1^{(2)}}{7 r_{12}^2}\nonumber\\
& + \frac{707 G^2 m_{1} m_{2} \mu_1^{(2)}}{8 r_{12}^2}
 + \frac{165 G^2 m_{2}^2 \mu_1^{(2)}}{4 r_{12}^2}\biggl]
 + \frac{15 \mu_1^{(3)}}{2 r_{12}^2}\biggl\}
 + 1 \leftrightarrow 2 + \mathcal{O}\left(
 \frac{\epsilon_\text{tidal}}{c^{6}}\right) \,.
\end{align}
Note that the last term, although it does not contain any explicit
$1/c$-factor, is actually a NNL term [see~\cref{eq:order}].

The long EoM derived by varying the Lagrangian~\eqref{eq:Ltidal} are relagated
to~\cref{appendix:accNNL}. We have verified that the latter EoM in harmonic
coordinates stay manifestly invariant when we perform a global (PN-expanded)
Lorentz boost with constant velocity $\bm{V}$. All the formulas employed to
check the Lorentz invariance are given by Eqs.~(3.20)--(3.23) of
Ref.~\cite{BFregM}. Furthermore, as a confirmation of the boost invariance of
the EoM, we can compute the Noetherian invariant associated with this
symmetry, which is nothing but the (mass weighted) position of the center of mass $G^i$
of the binary system. We obtain
$G^i = G^i_\text{pp} + G^i_\text{tidal}$, where the point-particle piece is
given by Eq.~(4.4) in~\cite{ABF01}, \textit{i.e.} at the 1PN order by
\begin{equation}\label{eq:Gipp}
G^i_\text{pp} = m_1 y_{1}^{i} + \frac{m_1}{2c^2}\left( v_1^2 -
  \frac{G m_2}{r_{12}} \right)y_{1}^{i} + 1\leftrightarrow2 +
\mathcal{O}\left( \frac{1}{c^{4}} \right)\,,
\end{equation}
and where the dominant tidal piece appears only at the NL/1PN order
and is given by
\begin{equation}\label{eq:Gitidal}
G^i_\text{tidal} = \frac{3 G^{2} m_{2}^{2}}{2 r_{12}^{5} c^2}
\,\mu_{1}^{(2)} \left(3 n_{12}^{i}- \frac{y_{1}^{i}}{r_{12}} \right) +
1\leftrightarrow2 + \mathcal{O}\left(
\frac{\epsilon_\text{tidal}}{c^{4}} \right)\,.
\end{equation}
For simplicity, since it is not needed in the following, we do not
present the complicated NNL/2PN contributions beyond the
result~\eqref{eq:Gitidal}.

\section{Tidal effects in the center-of-mass frame}
\label{sec:CoM}

The center-of-mass (CoM) frame is defined as the frame for which the
equation $G^i = 0$ holds, consistently including the tidal terms. The
structure of the leading order of the EoM and energy allows one to
compute the corresponding CoM quantities at the 2PN relative order
without requesting $G^{i}$ itself at that order. By contrast, it is
sufficient to know $G^{i}$ at the 1PN relative order for this
calculation, which means including the tidal effects at the NL/1PN
order as given by~\cref{eq:Gitidal}. Solving for $G^i = 0$ then yields
the CoM position of the particle 1 as a function of the relative
separation and velocity.\footnote{We pose
  $x^i=y_1^i-y_2^i$ and $v^i=\dd x^i/\dd t$;
  $r=\vert\bm{x}\vert=r_{12}$ denotes the separation, $n^i=x^i/r$ the unit
  direction, and we have
  $\dot{r}=(nv)=\bm{n}\cdot\bm{v}$; mass parameters are: the total mass
  $m=m_{1}+m_{2}$, the symmetric mass ratio $\nu = m_1 m_2/m^2 =
  X_{1}X_{2}$ and the mass difference $\Delta = X_{1}-X_{2}$, with
  $X_{A}=m_{A}/m$.} We find $y_1^i = (y_1^i)_\text{pp} +
(y_1^i)_\text{tidal}$, where the known 1PN expression for the point-particle piece reads
\begin{equation}\label{eq:y1pp}
(y_1^i)_\text{pp} = \left[ X_2 + \frac{\nu\,\Delta}{2 c^2} \left(v^2 - \frac{G
      m}{r}\right)\right] x^i + \mathcal{O}\left( \frac{1}{c^{4}} \right)\,,
\end{equation}
with the position of the particle 2 obtained by the exchange
$1\leftrightarrow 2$.  Now, the point is that, because of the tidal
contribution to the CoM position found in~\eqref{eq:Gitidal}, there
also exists a NL/1PN contribution given by
\begin{equation}\label{eq:y1tidal}
(y_1^i)_\text{tidal} = - \frac{3 G^{2} m \nu}{2 r^{6} c^2} \Big( \Delta
\,\mu_{+}^{(2)} +5 \mu_{-}^{(2)} \Big) x^{i}  + \mathcal{O}\left(
  \frac{\epsilon_\text{tidal}}{c^{4}} \right)\,.
\end{equation}
The velocities $v_1^i = (v_1^i)_\text{pp} + (v_1^i)_\text{tidal}$ are found by
iteratively differentiating Eqs.~\eqref{eq:y1pp}--\eqref{eq:y1tidal}, using in
that process the full EoM, which include the tidal effect. Here and below, we define
the following convenient combinations of the tidal polarizabilities:
\begin{equation}\label{eq:polarpm}
\mu_\pm^{(\ell)} = \frac{1}{2}\left(\frac{m_{2}}{m_{1}}\,\mu_{1}^{(\ell)} \pm
  \frac{m_{1}}{m_{2}}\,\mu_{2}^{(\ell)}\right)\,,\qquad \sigma_\pm^{(\ell)} =
\frac{1}{2}\left(\frac{m_{2}}{m_{1}}\,\sigma_{1}^{(\ell)} \pm
  \frac{m_{1}}{m_{2}}\,\sigma_{2}^{(\ell)}\right)\,,
\end{equation}
where the chosen normalisation is such that $\mu_+^{(\ell)} =
\mu_1^{(\ell)} = \mu_2^{(\ell)}$ and $\mu_-^{(\ell)} = 0$ when the two
bodies are identical, with the same mass and internal structure.
Likewise for $\sigma_\pm^{(\ell)}$.

At this stage, the EoM in the CoM frame can be derived in two possible ways:
either by computing the CoM acceleration $a^{i}= a_{1}^{i}-a_{2}^{i}$
directly, based on the replacement rules~\eqref{eq:y1pp}--\eqref{eq:y1tidal}, or
by getting first the expression of the Lagrangian in the CoM frame from the
Lagrangian in a general frame, varying it then to recover the EoM. We
resorted to the two methods and the results are in full
agreement (see also~\cite{MBBF17} for further details on the second method).
The CoM Lagrangian may be decomposed as $L = L_\text{pp} + L_\text{tidal}$,
where $L_\text{pp}$ is \textit{e.g.} given by Eq.~(4.2) in~\cite{BI03CM} while
the tidal part is, up to NNL order
\begin{align}\label{eq:LCoMtidal}
\frac{L_{\text{tidal}}}{\mu} &=\frac{G^2 m}{r^6} \Bigg\{ 3 \mu_{+}^{(2)}
 + \frac{1}{c^{2}} \biggl\{\biggl [\mu_{+}^{(2)} \Bigl(\frac{27}{2}
 + 9 \nu \Bigr)
 + \frac{45}{2} \Delta\mu_{-}^{(2)}
 - 24 \sigma_{+}^{(2)}\biggl] \dot{r}^2
 + \biggl [\mu_{+}^{(2)} \Bigl(\frac{15}{4}
 + \frac{3}{2} \nu \Bigr)
 -  \frac{15}{4} \Delta\mu_{-}^{(2)}\nonumber\\
& + 24 \sigma_{+}^{(2)}\biggl] v^{2}
 + \frac{G m}{r} \Bigl(- \frac{27}{2} \mu_{+}^{(2)}
 + \frac{15}{2} \Delta\mu_{-}^{(2)}\Bigr)\biggl\}
 + \frac{1}{c^{4}} \Biggl[r \biggl\{\biggl [\mu_{+}^{(2)} \Bigl(21
 -  \frac{45}{2} \nu \Bigr)
 + \Delta\mu_{-}^{(2)} \Bigl(21
 -  \frac{9}{2} \nu \Bigr)\nonumber\\
& - 48 \nu \sigma_{+}^{(2)}\biggl] a_{v} \dot{r}
 + \biggl [\mu_{+}^{(2)} \Bigl(-60
 + 18 \nu \Bigr)
 + \Delta\mu_{-}^{(2)} \Bigl(-60
 + 18 \nu \Bigr)
 + \sigma_{+}^{(2)} \Bigl(-16
 + 48 \nu \Bigr)
 - 16 \Delta\sigma_{-}^{(2)}\biggl] a_{n} \dot{r}^2\nonumber\\
& + \biggl [\mu_{+}^{(2)} \Bigl(\frac{39}{2}
 -  \frac{27}{4} \nu \Bigr)
 + \Delta\mu_{-}^{(2)} \Bigl(\frac{39}{2}
 -  \frac{9}{4} \nu \Bigr)
 + 16 \sigma_{+}^{(2)}
 + 16 \Delta\sigma_{-}^{(2)}\biggl] a_{n} v^{2}\biggl\}
 + \biggl [\mu_{+}^{(2)} \Bigl(36
 - 72 \nu
 + 18 \nu^2\Bigr)\nonumber\\
& + \Delta\mu_{-}^{(2)} \Bigl(27
 - 18 \nu \Bigr)
 + \sigma_{+}^{(2)} \Bigl(72
 - 96 \nu \Bigr)
 + 48 \Delta\sigma_{-}^{(2)}\biggl] \dot{r}^4
 + \biggl [\mu_{+}^{(2)} \Bigl(- \frac{189}{4}
 + 72 \nu
 -  \frac{45}{2} \nu^2\Bigr)
 + \Delta\mu_{-}^{(2)} \Bigl(- \frac{99}{4}\nonumber\\
& -  \frac{27}{2} \nu \Bigr)
 + \sigma_{+}^{(2)} \Bigl(-114
 + 132 \nu \Bigr)
 - 54 \Delta\sigma_{-}^{(2)}\biggl] \dot{r}^2 v^{2}
 + \biggl [\mu_{+}^{(2)} \Bigl(\frac{249}{16}
 - 12 \nu
 -  \frac{27}{8} \nu^2\Bigr)
 + \Delta\mu_{-}^{(2)} \Bigl(\frac{39}{16}
 + \frac{27}{8} \nu \Bigr)\nonumber\\
& + \sigma_{+}^{(2)} \Bigl(42
 - 36 \nu \Bigr)
 + 6 \Delta\sigma_{-}^{(2)}\biggl] v^{4}
 + \frac{G m}{r} \biggl\{\biggl [\mu_{+}^{(2)} \Bigl(- \frac{249}{2}
 + \frac{355}{2} \nu
 + 39 \nu^2\Bigr)
 + \Delta\mu_{-}^{(2)} \Bigl(- \frac{303}{2}
 + \frac{135}{2} \nu \Bigr)\nonumber\\
& + 28 \sigma_{+}^{(2)}
 - 44 \Delta\sigma_{-}^{(2)}\biggl] \dot{r}^2
 + \biggl [\mu_{+}^{(2)} \Bigl(\frac{123}{4}
 - 41 \nu
 + 3 \nu^2\Bigr)
 + \frac{213}{4} \Delta\mu_{-}^{(2)}
 - 28 \sigma_{+}^{(2)}
 + 44 \Delta\sigma_{-}^{(2)}\biggl] v^{2}\biggl\}\nonumber\\
& + \frac{G^2 m^2}{r^2} \biggl [\mu_{+}^{(2)} \Bigl(\frac{915}{28}
 + \frac{3119}{28} \nu \Bigr)
 -  \frac{1395}{28} \Delta\mu_{-}^{(2)}\biggl]\Biggl]
 + \mu_{+}^{(3)} \frac{15}{r^2}\Bigg\} + \mathcal{O}\left(
  \frac{\epsilon_\text{tidal}}{c^{6}} \right)\,.
\end{align}
Note again that the last term is actually a NNL/2PN contribution. The
corresponding relative CoM acceleration is displayed
in~\cref{appendix:accNNL}. Similarly, we show here the tidal part of
the conserved energy $E = E_\text{pp} + E_\text{tidal}$:
\begin{align}\label{eq:Etidal}
  \frac{E_\text{tidal}}{m\nu} =& -
  3 \frac{G^{2} m }{r^{6}}\mu_{+}^{(2)} +
  \frac{1}{c^{2}} \left\{ \frac{G^{2} m}{r^{6}}
  \left[ \left[ \left(\frac{27}{2} + 9\nu \right)\mu_{+}^{(2)} +
 \frac{45}{2} \Delta \, \mu_{-}^{(2)} -24 \sigma_{+}^{(2)} \right]\dot{r}^{2}
 \right. \right. \nonumber\\
 & \left. \left. + \left( \left(\frac{15}{4} +
   \frac{3}{2}\nu \right)\mu_{+}^{(2)} -
   \frac{15}{4} \Delta \, \mu_{-}^{(2)} +
   24 \sigma_{+}^{(2)} \right) v^{2} \right] +
   \frac{G^{3} m^{2}}{r^{7}} \left[ \frac{27}{2} \mu_{+}^{(2)} -
   \frac{15}{2} \Delta \, \mu_{-}^{(2)} \right] \right\} \nonumber \\
 & + \frac{1}{c^{4}} \left\{ \frac{G^{2} m}{r^{6}}
   \left[ \left( \left( -372-72 \nu +54 \nu^{2} \right)\mu_{+}^{(2)} +
   \left(-399 +90 \nu \right) \Delta \, \mu_{-}^{(2)} +
   \left(88+96\nu \right)\sigma_{+}^{(2)} +
   16 \Delta \, \sigma_{-}^{(2)}\right)\dot{r}^{4} \right. \right. \nonumber \\
 & \left. \left. + \left( \left( \frac{1125}{4}-
   \frac{27}{2} \nu -\frac{135}{2} \nu^{2} \right)\mu_{+}^{(2)} +
   \left(\frac{1395}{4} -135 \nu \right) \Delta \, \mu_{-}^{(2)} +
   \left(-198-36\nu \right)\sigma_{+}^{(2)} -18 \Delta \, \sigma_{-}^{(2)}
   \right)
   \dot{r}^{2}v^{2} \right. \right. \nonumber \\
 & \left. \left. + \left( \left( \frac{99}{16}-
   \frac{27}{4} \nu -\frac{81}{8} \nu^{2} \right)\mu_{+}^{(2)} +
   \left(-\frac{531}{16} +\frac{135}{8} \nu \right)
   \Delta \, \mu_{-}^{(2)} + \left(110-60\nu \right)\sigma_{+}^{(2)} +
   2 \Delta \, \sigma_{-}^{(2)}\right)v^{4}\right] \right. \nonumber \\
 & \left. + \frac{G^{3} m^{2}}{r^{7}}
   \left[ \left( \left( -\frac{213}{2}+\frac{499}{2} \nu +
   39 \nu^{2} \right)\mu_{+}^{(2)} +
   \left(-\frac{267}{2} +\frac{135}{2} \nu \right) \Delta \, \mu_{-}^{(2)} +
   \left(60+48\nu \right)\sigma_{+}^{(2)} -
   12 \Delta \, \sigma_{-}^{(2)}\right) \dot{r}^{2} \right. \right. \nonumber \\
 & \left. \left. + \left( \left( \frac{51}{4}-
  113 \nu +3 \nu^{2} \right)\mu_{+}^{(2)} +\frac{141}{4}  \Delta \, \mu_{-}^{(2)} +
   \left(-60-48\nu \right)\sigma_{+}^{(2)} +
   12 \Delta \, \sigma_{-}^{(2)}\right) v^{2} \right] \right. \nonumber \\
 & \left. + \frac{G^{4} m^{3}}{r^{8}}
   \left[ \left( -\frac{915}{28} - \frac{3119}{28} \nu \right)\mu_{+}^{(2)} +
   \frac{1395}{28}\Delta  \, \mu_{-}^{(2)}\right] \right\} -
   15 \frac{G^{2} m }{r^{8}}\mu_{+}^{(3)} + \mathcal{O}\left(
  \frac{\epsilon_\text{tidal}}{c^{6}} \right)\,.
\end{align}
Finally, for the CoM angular momentum
$J^{i} = J^{i}_\text{pp} + J^{i}_\text{tidal}$, we find (denoting
$L^i=\varepsilon_{ijk}x^{j}v^{k}$)
\begin{align}\label{eq:Jtidal}
\frac{J^{i}_\text{tidal}}{m\nu} &= \frac{G^2 m}{c^2 r^6} L^{i}
 \Biggl[\mu_{+}^{(2)} \Bigl(\frac{15}{2} 
 + 3 \nu \Bigr)
 -  \frac{15}{2} \Delta\mu_{-}^{(2)}
 + 48 \sigma_{+}^{(2)}
 + \frac{1}{c^{2}} \biggl\{\biggl [\mu_{+}^{(2)} \Bigl(\frac{303}{2}
 - 27 \nu
 - 45 \nu^2\Bigr)
 + \Delta\mu_{-}^{(2)} \Bigl(\frac{393}{2}
 - 90 \nu \Bigr)\nonumber\\
& + \sigma_{+}^{(2)} \Bigl(-196
 - 120 \nu \Bigr)
 - 76 \Delta\sigma_{-}^{(2)}\biggl] \dot{r}^2
 + \biggl [\mu_{+}^{(2)} \Bigl(\frac{9}{4}
 - 12 \nu
 -  \frac{27}{2} \nu^2\Bigr)
 + \Delta\mu_{-}^{(2)} \Bigl(- \frac{201}{4}
 + \frac{45}{2} \nu \Bigr)
 + \sigma_{+}^{(2)} \Bigl(136\nonumber\\
& - 96 \nu \Bigr)
 - 8 \Delta\sigma_{-}^{(2)}\biggl] v^{2}
 + \frac{G m}{r} \biggl [\mu_{+}^{(2)} \Bigl(\frac{87}{2}
 - 154 \nu
 + 6 \nu^2\Bigr)
 + \frac{177}{2} \Delta\mu_{-}^{(2)}
 + \sigma_{+}^{(2)} \Bigl(-88
 - 48 \nu \Bigr)
 + 56 \Delta\sigma_{-}^{(2)}\biggl]\biggl\}\Biggl]\nonumber \\ & + \mathcal{O}\left(
  \frac{\epsilon_\text{tidal}}{c^{6}} \right)\,.
\end{align}
The point-particle pieces $E_{\text{pp}}$ and $J^i_{\text{pp}}$ are depicted
in Eqs.~(4.8) and~(4.9) of Ref.~\cite{BI03CM}.

\section{Tidal effects for quasi-circular orbits}
\label{sec:circ}

We consider quasi-circular orbits, \textit{i.e.} orbits that are
circular in our harmonic coordinate system but for the dissipative
radiation-reaction effects. For such orbits, we can neglect
$\dot{r}=\mathcal{O}(c^{-5})$, which is precisely of the order of radiation
reaction effects. Under this assumption, we see from~\cref{eq:accCoM} that the CoM
acceleration becomes purely radial, $a^{i} = - \omega^{2} x^{i}$, from
which we can read off the orbital angular frequency $\omega$. Relevant
quantities will then depend only on the bodies' separation $r$ or,
equivalently (\textit{via} a generalized Kepler third law), on the orbital
frequency $\omega$. In the case of circular orbits, it is convenient to introduce
the dimensionless PN parameters associated with the separation and
orbital frequency as
\begin{equation}\label{eq:defgammax}
\gamma = \frac{Gm}{rc^2}\,,\qquad x=\left( \frac{G m \omega}{c^3}\right)^{2/3}\,,
\end{equation}
as well as to adimensionalize the polarizability coefficients defined
in Eqs.~\eqref{eq:polarpm} by considering the ``tilded''
quantities\footnote{The quantity $\kappa^{T}_{2}$ defined in
  Ref.~\cite{DNV12} is related to our definition
  $\widetilde{\mu}_{+}^{(2)}$ by $\kappa^{T}_{2} = 6
  \widetilde{\mu}_{+}^{(2)}$.}
\begin{equation}\label{eq:polarpmtilde}
\widetilde{\mu}_\pm^{(\ell)} = \left(\frac{c^2}{G m}\right)^{2\ell+1}
\!\!\!G\,\mu_\pm^{(\ell)}\,,\qquad \widetilde{\sigma}_\pm^{(\ell)} =
\left(\frac{c^2}{G m}\right)^{2\ell+1} \!\!\!G\,\sigma_\pm^{(\ell)}\,.
\end{equation}
By identifying the expression of $\omega^{2}$ from the circular-orbit
EoM as explained above and replacing $\gamma$ iteratively, we recover
the well-known formula for point masses at the 2PN order, with
a non-trivial NNL/2PN relative tidal contribution
\begin{subequations}\label{eq:omega2}
\begin{align}
(\omega^{2})_\text{pp} &= \frac{G m}{r^{3}}\left[ 1 + (-3+\nu)\gamma +
    \left( 6 + \frac{61}{4} \nu + \nu^{2} \right) \gamma^{2}\right] +
  \mathcal{O}\left( \frac{1}{c^{6}} \right)
  \,,\\ (\omega^{2})_\text{tidal} &= \frac{G m}{r^{3}}\biggl\{
  18\,\widetilde{\mu}_{+}^{(2)} \gamma^{5} + \left[ \left(
    -\frac{249}{2} +51 \nu \right)\widetilde{\mu}_{+}^{(2)} +
    \frac{75}{2}\Delta \, \widetilde{\mu}_{-}^{(2)}
    +96\,\widetilde{\sigma}_{+}^{(2)} \right] \gamma^{6}
  \nonumber\\ &\qquad\quad + \left[ \left( \frac{34317}{56} +
    \frac{2976}{7} \nu + 54\nu^{2}\right)\widetilde{\mu}_{+}^{(2)} +
    \left( -\frac{12051}{56} +90\nu \right) \Delta \,
    \widetilde{\mu}_{-}^{(2)} \right. \nonumber\\& \qquad\qquad\qquad
    \left. + \bigl(-616+264\nu\bigr)\widetilde{\sigma}_{+}^{(2)} + 200
    \Delta \, \widetilde{\sigma}_{-}^{(2)} + 120
    \,\widetilde{\mu}_{+}^{(3)}\right]\gamma^{7}\biggr\} +
  \mathcal{O}\left( \frac{\epsilon_\text{tidal}}{c^{6}} \right)\,.
\end{align}\end{subequations}
Next, we may determine the relation between $\gamma$ and $x$, defined 
in Eqs.~\eqref{eq:defgammax}, by inverting
Eqs.~\eqref{eq:omega2}, with result:
\begin{subequations}\label{eq:gammaofx}
\begin{align}
\gamma_\text{pp} &= x \left[ 1 + \left(1-\frac{\nu}{3}\right)x +
  \left( 1 - \frac{65}{12} \nu \right)x^{2} \right] +
\mathcal{O}\left(\frac{1}{c^{6}} \right) \,,\\ \gamma_\text{tidal} &=
x \biggl\{ -6\widetilde{\mu}_{+}^{(2)} x^{5} + \left[ \left(
  -\frac{37}{2} +3 \nu \right)\widetilde{\mu}_{+}^{(2)} -
  \frac{25}{2}\Delta \, \widetilde{\mu}_{-}^{(2)}
  -32\widetilde{\sigma}_{+}^{(2)} \right] x^{6}
\nonumber\\ &\qquad\quad + \left[ \left( -\frac{4355}{56} +
  \frac{1105}{21} \nu + 15\nu^{2}\right)\widetilde{\mu}_{+}^{(2)} +
  \left( -\frac{3683}{56} +\frac{95}{6}\nu \right) \Delta \,
  \widetilde{\mu}_{-}^{(2)} \right. \nonumber\\& \qquad\qquad\qquad
  \left. +
  \left(-\frac{440}{3}+\frac{88}{3}\nu\right)\widetilde{\sigma}_{+}^{(2)}
  - \frac{200}{3} \Delta \, \widetilde{\sigma}_{-}^{(2)} - 40
  \widetilde{\mu}_{+}^{(3)}\right]x^{7}\biggr\} + \mathcal{O}\left(
\frac{\epsilon_\text{tidal}}{c^{6}} \right)\,.
\end{align}\end{subequations}
The conserved energy for circular orbits can now be computed. To do
so, we take \cref{eq:Etidal} to which we add the point-particle part,
set $\dot{r}=0$ and replace $v^{2}=r^{2}\omega^{2}$ by its expression in terms of
the parameter $\gamma$ using Eqs.~\eqref{eq:omega2}. This yields $E$
first as a function of $\gamma$. We finally insert there the previous
relation~\eqref{eq:gammaofx} between $\gamma$ and $x$ to get an
important result, namely the expression of the circular
energy as a function of the frequency-dependent parameter $x$:
\begin{subequations}\label{eq:Ex}
\begin{align}
E_\text{pp} &= -\frac{1}{2} m \nu x c^2 \left[ 1 + \left( -\frac{3}{4}
                -\frac{\nu}{12} \right)x + \left( -\frac{27}{8} +\frac{19}{8}
                \nu - \frac{\nu^{2}}{24} \right)x^{2} \right] +
              \mathcal{O}\left(\frac{1}{c^{6}} \right)\,, \\
E_\text{tidal} &= -\frac{1}{2} m \nu x c^2
    \biggl\{ - 18 \widetilde{\mu}_{+}^{(2)} x^5 +
    \left[\left(-\frac{121}{2} +33\nu \right)\widetilde{\mu}_{+}^{(2)}-
    \frac{55}{2}\Delta \, \widetilde{\mu}_{-}^{(2)} -
    176\,\widetilde{\sigma}_{+}^{(2)} \right]x^6 \nonumber\\ 
  &\qquad\quad + \left[ \left(-\frac{20865}{56} +
    \frac{5434}{21}\nu -\frac{91}{4}\nu^2 \right)
    \widetilde{\mu}_{+}^{(2)} +
    \Delta \left(-\frac{11583}{56}+\frac{715}{12}\nu \right)
    \widetilde{\mu}_{-}^{(2)} \right. \nonumber\\& \qquad\qquad\qquad \left. +
  \left(-\frac{2444}{3} +
    \frac{1768}{3}\nu \right)\widetilde{\sigma}_{+}^{(2)} -
    \frac{884}{3} \Delta \, \widetilde{\sigma}_{-}^{(2)} -
    130 \,\widetilde{\mu}_{+}^{(3)} \right]x^7 \biggr\} + \mathcal{O}\left(
  \frac{\epsilon_\text{tidal}}{c^{6}} \right)\,.
\end{align}
\end{subequations}
We can also compute by the same method the constant angular momentum
for circular orbits, which reads
\begin{subequations}\label{eq:Jx}
\begin{align}
J_\text{pp} &= \frac{G m^2 \,\nu}{c\,x^{1/2}} \left[ 1 + \left(
    \frac{3}{2} + \frac{\nu}{6} \right) x + \left( \frac{27}{8} -
    \frac{19}{8} \nu + \frac{\nu^2}{24} \right) x^2\right] +
              \mathcal{O}\left(\frac{1}{c^{6}} \right)\,,\\
J_\text{tidal} &= \dfrac{G m^{2} \nu}{c x^{1/2}} \biggl\{12
\widetilde{\mu}_{+}^{(2)} x^5 +
\left[\left(\dfrac{77}{2} -21\nu
  \right)\widetilde{\mu}_{+}^{(2)}+\dfrac{35}{2}\Delta \,
  \widetilde{\mu}_{-}^{(2)} +112 \widetilde{\sigma}_{+}^{(2)}
  \right]x^6 + \left[ \left(\dfrac{1605}{7} - \dfrac{3344}{21}\nu
  +14\nu^2 \right)\widetilde{\mu}_{+}^{(2)} \right. \nonumber \\ &
  \qquad \qquad \left. + \Delta \left(\dfrac{891}{7}-\dfrac{110}{3}\nu
  \right)\widetilde{\mu}_{-}^{(2)} + \left(\dfrac{1504}{3} -
  \dfrac{1088}{3}\nu \right)\widetilde{\sigma}_{+}^{(2)} +
  \dfrac{544}{3} \Delta \, \widetilde{\sigma}_{-}^{(2)} +80
  \widetilde{\mu}_{+}^{(3)} \right]x^7 \biggr\} + \mathcal{O}\left(
\frac{\epsilon_\text{tidal}}{c^{6}} \right)\,.
\end{align}
\end{subequations}
We have verified that the energy $E$ and angular momentum $J$ for
circular orbits, including all the tidal contributions given
in~\eqref{eq:Ex}--\eqref{eq:Jx}, are linked by the famous
relation
\begin{equation}\label{eq:thermo}
\frac{\partial E}{\partial \omega} = \omega \, \frac{\partial
  \mathrm{J}}{\partial \omega} + \mathcal{O}\left( \frac{1}{c^{6}},
  \frac{\epsilon_\text{tidal}}{c^{6}}\right)\,,
\end{equation}
which is just one aspect of the ``first law of binary point-particle
mechanics''~\cite{LBW12}.

\section{Summary and conclusions}\label{sec:conclusion}

We have computed the Lagrangian and associated conserved quantities of compact
binaries including tidal interactions up to the NNL order, corresponding to
the 2PN approximation beyond the leading quadrupolar tidal effect occuring at
the 5PN order. The results follow from the effective Fokker
action~\eqref{eq:Sg}, \eqref{eq:Sm} with non-minimal matter couplings, and
are parametrized by polarizability coefficients describing the mass
quadrupole, mass octupole and current quadrupole tidal interactions. In
particular, we have obtained the NNL conserved invariant energy of the compact
binary for quasi-circular orbits.

To conclude, let us compare our expressions for the invariant energy as given
by~\eqref{eq:Ex} with existing results in the literature. In the following
table, we provide for each order and for each multipolar piece contributing to
the conserved energy $E_\text{tidal}(x)$ the references which we agree with:
\begin{center}
\begin{tabular}{|c||c|c|c|}
\hline $E_\text{tidal}$ & Mass quadrupole & Current quadrupole & Mass
octupole \\ \hline \hline 5PN (L) & \cite{FH08, VHF11, BiniDF12, VF13,
  F14} & $\times$ & $\times$ \\ \hline 6PN (NL) & \cite{BiniDF12,
  VF13, AGP18}& \cite{BiniDF12, AGP18, BV18} & $\times$ \\ \hline 7PN
(NNL) & \cite{BiniDF12} & \cite{BiniDF12} & \cite{BiniDF12, Landry18}
\\ \hline
\end{tabular}
\end{center}
Note in particular that we are in full agreement with all results of
Ref.~\cite{BiniDF12}. We have checked, notably, that by re-expanding the tidal
effects entering the EOB Hamiltonian~\cite{BiniDF12} in the form of a PN
Taylor series, we recover exactly our
equation~\eqref{eq:Ex}.\footnote{However, we do not recover the 1PN
  coefficient for the current quadrupole piece in Ref.~\cite{Landry18}, where
  the discrepancy is by a factor 2.}

Now that the problem of the Lagrangian and EoM is solved
(Ref.~\cite{BiniDF12} and this work), we shall compute in a second
paper~\cite{article_flux} the gravitational-wave energy flux for
quasi-circular orbits, and then, from it, deduce, through the energy balance
equation, the crucial orbital phase and frequency evolution (or
``chirp'') of compact binaries in circular orbits including tidal
effects up to the NNL/2PN order beyond the Einstein quadrupole
formula.

\acknowledgments

We thank Gilles Esposito-Far\`ese for useful discussions. We are also greatful to
Justin Vines for interesting informative discussions during the preparation of
this work.

\appendix

\section{Newtonian treatment of the tidal effects}
\label{appendix:Newtonian}

In this Appendix, we derive the Newtonian EoM and the Lagrangian of a
system of $N$ extended compact bodies without spins, including
multipolar tidal interaction effects. The mass and the CoM position of
each of the objects are defined by
\begin{equation}\label{eq:mAyACM}
m_A = \int_{\mathcal{V}_A} \dd^3\mathbf{x} \,\rho(\mathbf{x},t)\,,\qquad y_A^i(t) =
\frac{1}{m_A} \int_{\mathcal{V}_A} \dd^3\mathbf{x} \,\rho(\mathbf{x},t) \,x^i\,,
\end{equation}
where the integrals extend over the volume $\mathcal{V}_A$ of body
$A$, and where $\rho(\mathbf{x},t)$ denotes the Eulerian density of
the $N$-body system satisfying the usual continuity equation
$\partial_t\rho+\partial_i(\rho v^i)=0$ (hence the mass $m_A$ is
constant). The equation of motion verified by the CoM line of body $A$
is then given by
\begin{equation}\label{eq:accA}
m_A \frac{\dd^2 y_A^i}{\dd t^2} = \sum_{B\not= A} \int_{\mathcal{V}_A} \dd^3\mathbf{x}
\,\rho\,\partial_i U_B\,,
\end{equation}
where we have discarded the self-field of body $A$ which is zero by Newton's
action-reaction theorem (so the sum runs over all the bodies $B\not=A$), and
where the Newtonian potential generated by body $B$ reads
\begin{equation}\label{eq:UB}
U_B(\mathbf{x},t) = G \int_{\mathcal{V}_B}
\frac{\dd^3\mathbf{x}'}{\vert\mathbf{x}-\mathbf{x}'\vert}
\,\rho(\mathbf{x}',t)\,.
\end{equation}
For any point outside the body $B$, thus in particular located inside
the body $A$, distinct from $B$, we have $\Delta U_B=0$. Next, we
define the Newtonian STF multipole moment of body $A$ to be
\begin{equation}\label{eq:IAL}
I_A^{L}(t) = \int_{\mathcal{V}_A} \dd^3\bm{z}_A
\,\rho_A(\bm{z}_A,t)\,\hat{z}_A^L\,,
\end{equation}
where we adopted as integration variable the distance
$\bm{z}_A=\mathbf{x}-\bm{y}_A(t)$ linking the line of the CoM
$\bm{y}_A(t)$ to the generic point $\mathbf{x}\in\mathcal{V}_A$, where
$\hat{z}_A^L=\text{STF}(z_A^L)$ denotes the STF product of $\ell$
spatial vectors $z_A^L=z_A^{i_1}\cdots z_A^{i_\ell}$ (with
$L=i_1\cdots i_\ell$ a multi-spatial index), and where we have posed
$\rho_A(\bm{z}_A,t)=\rho(\bm{y}_a+\bm{z}_a,t)$. With this notation the
mass monopole moment is just the constant mass, while the CoM position
$y_A^i$ is defined by the nullity of the mass dipole moment:
\begin{equation}\label{eq:monodipole}
I_A = m_A\,,\qquad I_A^i=0\,.
\end{equation}

On the other hand, the Newtonian tidal moments, starting with the
quadrupole moment ($\ell\geqslant 2$), are defined quite naturally as
the multi-gradients of the total external potential due to the other
bodies felt by the body $A$ at the location of its CoM $\bm{y}_A$:
\begin{equation}\label{eq:GAL}
G_A^{L}(t) = \sum_{B\not= A} \bigl(\partial_L
U_B\bigr)(\bm{y}_A)\qquad\text{(for $\ell\geqslant 2$)}\,,
\end{equation}
with $\partial_L=\partial_{i_1}\cdots\partial_{i_\ell}$. Since $\Delta
U_B=0$ inside body $A$, the tidal moments are automatically STF in all
their indices $L$, namely $\partial_L U_B=\hat{\partial}_L U_B$. For
the dipolar tidal moment (with $\ell=1$) it is convenient to pose
\begin{equation}\label{eq:GAi}
G_A^{i} = \sum_{B\not= A} \bigl(\partial_i U_B\bigr)(\bm{y}_A) - \frac{\dd^2
  y_A^i}{\dd t^2}\,,
\end{equation}
so that $G_A^{i}=0$ for a system of point particles described only by
their masses, their higher multipole moments being neglected. The EoM
may then be rewritten in elegant form as (see
\textit{e.g.}~\cite{DSX1})
\begin{equation}\label{eq:EOM}
m_A\,G_A^{i} + \sum_{\ell=2}^{+\infty} \frac{1}{\ell!} \,I_A^L\,G_A^{iL} = 0\,.
\end{equation}
Using the fact that for any $\mathbf{x}$ outside the body $B$ we have
the multipole decomposition
\begin{equation}\label{eq:UBexp}
U_B = G \sum_{k=0}^{+\infty} \frac{(-)^k}{k!}
\,I_B^K\,\partial_{K}\Bigl(\frac{1}{r_B}\Bigr)\,,
\end{equation}
with $r_{B} = \vert\mathbf{x}-\bm{y}_B\vert$, we see that the tidal
moments themselves~\eqref{eq:GAL} can be expanded in terms of the
multipole moments of the other bodies as (for $\ell\geqslant 2$)
\begin{equation}\label{eq:GALexp}
G_A^{L} = G \sum_{B\not= A} \,\sum_{k=0}^{+\infty} \frac{(-)^k}{k!}
\,I_B^K\,\partial_{LK}^A\Bigl(\frac{1}{r_{AB}}\Bigr)\,,
\end{equation}
where $r_{AB} = \vert\bm{y}_A-\bm{y}_B\vert$ is the distance between the CoMs
of the bodies $A$ and $B$, the gradient is taken with respect to the point
$A$, \textit{i.e.} $\partial_i^A=\partial/\partial y_A^i$, and we denote
$\partial_{LK}^A=\partial_{L}^A\partial_{K}^A$ with
$\partial_{L}^A=\partial_{i_1}^A\cdots\partial_{i_\ell}^A$. Finally, the EoM
admit the double multipole expansion series
\begin{equation}\label{eq:EOMexp}
m_A \frac{\dd^2 y_A^i}{\dd t^2} = G \sum_{B\not= A}
\,\sum_{\ell=0}^{+\infty}\,\sum_{k=0}^{+\infty} \frac{(-)^k}{\ell! \,k!}
\,I_A^L\,I_B^K\,\partial_{iLK}^A\Bigl(\frac{1}{r_{AB}}\Bigr)\,,
\end{equation}
or in more details (see \textit{e.g.} Eq.~(1.201)
of~\cite{PoissonWill}),
\begin{align}\label{eq:EOMdetail}
  m_A \frac{\dd^2 y_A^i}{\dd t^2} = G \sum_{B \not= A} \biggl\{ m_A
  m_B \,\partial_i^A\Bigl(\frac{1}{r_{AB}}\Bigr) & +
  \sum_{\ell=2}^{+\infty}\frac{(-)^\ell}{\ell!}\Bigl[ m_A \,I_B^{L} +
    (-)^\ell m_B \,I_A^{L}
    \Bigr]\partial_{iL}^A\Bigl(\frac{1}{r_{AB}}\Bigr) \nonumber \\ & +
  \sum_{\ell=2}^{+\infty}\,\sum_{k=2}^{+\infty} \frac{(-)^k}{\ell!
    \,k!}
  \,I_A^L\,I_B^K\,\partial_{iLK}^A\Bigl(\frac{1}{r_{AB}}\Bigr)\biggr\}
  \,.
\end{align}
Those equations have been generalized to the 1PN order~\cite{Xu97,
  Wu98, Racine05, VF13} using the DSX formalism~\cite{DSX1, DSX2}.

We now consider the case where the multipole moments are exclusively
induced by the tidal field of the other bodies. To describe this
situation, we assume that each extended body is at hydrodynamical
equilibrium at every time, so that the mass distribution at any
instant is aligned on the equipotentials of the external gravitational
field. We are thus in the so-called adiabatic regime where the
relaxation time scale of the body internal dynamics is significantly
smaller than the orbital time scale. In particular, we neglect the
dissipative effects due to the tides, considering only the
conservative dynamics of the system, and look for a Lagrangian. In
this case, we introduce a linear-response coefficient $\mu^{(\ell)}$
depending on the internal structure of the body and characterizing its
deformability or ``polarizability'' under the influence of the
external field, such that its multipole moments obey
\begin{equation}\label{eq:muAdef}
I_A^L = \mu_A^{(\ell)} \,G_A^L\,.
\end{equation}
Following usual definitions (see
\textit{e.g.}~\cite{Hind08,BinnP09,DN09tidal}), this coefficient is
related to the radius $R$ of the body and the (mass-type) multipolar
Love numbers $k^{(\ell)}$ by
\begin{equation}\label{eq:relLove}
G\,\mu_A^{(\ell)} = \frac{2}{(2\ell-1)!!} \,k_A^{(\ell)}\,R_A^{2\ell+1}\,.
\end{equation}
The Newtonian EoM~\eqref{eq:EOMdetail} become now
\begin{align}\label{eq:EOMpolar}
  m_A \frac{\dd^2 y_A^i}{\dd t^2} = G \sum_{B \not= A} \biggl\{ m_A
  m_B \,\partial_i^A\Bigl(\frac{1}{r_{AB}}\Bigr) &+
  \sum_{\ell=2}^{+\infty}\frac{(-)^\ell}{\ell!}  \Bigl[ m_A
    \,\mu_B^{(\ell)}\,G_B^{L} + (-)^\ell m_B \,
    \mu_A^{(\ell)}\,G_A^{L}
    \Bigr]\partial_{iL}^A\Bigl(\frac{1}{r_{AB}}\Bigr)
  \nonumber\\ \qquad\qquad &+
  \sum_{\ell=2}^{+\infty}\,\sum_{k=2}^{+\infty} \frac{(-)^k}{\ell!
    \,k!}  \,\mu_A^{(\ell)}\mu_B^{(k)}\,G_A^L\,G_B^K\,
  \partial_{iLK}^A\Bigl(\frac{1}{r_{AB}}\Bigr)\biggr\} \,,
\end{align}
in which the tidal moments obey the implicit relation
\begin{equation}\label{eq:GLAexpl}
G_A^{L} = G \sum_{B\not= A} \biggl[
  m_B\,\partial_{L}^A\Bigl(\frac{1}{r_{AB}}\Bigr) +
  \sum_{k=2}^{+\infty} \frac{(-)^k}{k!}
  \,\mu_B^{(k)}\,G_B^K\,\partial_{LK}^A\Bigl(\frac{1}{r_{AB}}\Bigr)\biggr]\,.
\end{equation}
The latter equations describe the conservative dynamics of the system
of $N$ extended bodies. The dependence on the internal structure is
entirely carried out by the coefficients $\mu^{(\ell)}$, which are
supposed to be constant. The dynamics is conservative in the sense
that it can be derived from the following exact Lagrangian, valid up
to any order in the multipole expansion and the tidal moments:
\begin{align}\label{eq:LN}
  L &= \sum_A \biggl\{ \frac{1}{2}m_A v_A^2 +\frac{1}{2}m_A\sum_{B\neq
    A} U_B(\bm{y}_A) \biggr\} \nonumber \\ &= \sum_A \biggl\{
  \frac{1}{2}m_A v_A^2 + \frac{1}{2}\sum_{\ell=2}^{+\infty}
  \frac{1}{\ell!}\, \mu_A^{(\ell)}\,G_A^{L}\,G_A^{L} + G \sum_{B > A}
  \biggl[\frac{ m_A m_B}{r_{AB}} -
    \sum_{\ell=2}^{+\infty}\,\sum_{k=2}^{+\infty} \frac{(-)^k}{\ell!
      \,k!}  \,\mu_A^{(\ell)}\,\mu_B^{(k)}\,G_A^L\,G_B^K\,
    \partial_{LK}^A\Bigl(\frac{1}{r_{AB}}\Bigr)\biggr] \biggr\}\,.
\end{align}
The Newtonian action is formally the Newtonian limit, at the quadratic
level, of the non-minimal matter action~\eqref{eq:Sm} in general
relativity. However, the action~\eqref{eq:Sm} is effective (or
``skeletonized''), with each compact object described by an effective
point particle endowed with internal structure. The mass-type moments
$G^{\hat{L}}$ (even parity sector) entering~\cref{eq:Sm} tend towards
the Newtonian tidal moments $G^{L}$, so that they can be regarded as their
legitimate relativistic versions, and the corresponding response
coefficients $\mu^{(\ell)}$ identify with the Newtonian tidal
deformabilities. Moreover, the relativistic action also depends on
current-type moments $H^{\hat{L}}$ (odd parity sector) with associated
response coefficients $\sigma^{(\ell)}$, first arising at the 1PN
relativistic order.

Both sets of relativistic tidal moments are given by appropriate
covariant derivatives of the Riemann tensor, which is nothing but the
relativistic tidal field felt by the body. Those moments are evaluated
at the location of the particle and a UV-type regularization is
required to remove the self field of that particle. Thus, in the
effective action, the self-field regularization automatically selects
the external tidal field experienced by the body due to the other
bodies composing the system.

\section{Proof that the trace terms to NNL order can be removed by a
  redefinition of the metric} 
\label{appendix:proof}

In this section, we show that the tidal moments entering the action
may be defined in terms of the Riemann tensor instead of the Weyl
tensor, since the traces of the Riemann tensor do not play any role in
the dynamics. Here, we shall denote $G_{\mu\nu}^{(R)}$,
$G_{\lambda\mu\nu}^{(R)}$ and $H_{\mu\nu}^{(R)}$ the tidal
mass-quadrupole, mass-octupole and current-quadrupole moments
introduced in Eqs.~\eqref{eq:tidalR}, while $G_{\mu\nu}^{(C)}$,
$G_{\lambda\mu\nu}^{(C)}$ and $H_{\mu\nu}^{(C)}$ will represent the
same but built with the Weyl tensor instead of the Riemann tensor. We
thus pose (setting $G=c=1$, and omitting particles' labels and mention
of the regularization)
\begin{subequations}\label{eq:tidalRC}
\begin{align}
&G^{(R)}_{\mu\nu} = - R_{\mu\rho\nu\sigma} u^{\rho}u^{\sigma}\,,
  &G^{(C)}_{\mu\nu} = - C_{\mu\rho\nu\sigma} u^{\rho}u^{\sigma}\,,
  \\ &H^{(R)}_{\mu\nu} = 2 R^{*}_{(\mu\underline{\rho}\nu)\sigma}
  u^{\rho}u^{\sigma}\,, &H^{(C)}_{\mu\nu} =2
  C^{*}_{(\mu\underline{\rho}\nu)\sigma}
  u^{\rho}u^{\sigma}\,,\\ &G^{(R)}_{\lambda\mu\nu} = -
  \nabla^\perp_{(\lambda}
  \,R_{\mu\underline{\rho}\nu)\sigma} u^{\rho}
  u^{\sigma}\,, &G^{(C)}_{\lambda\mu\nu} = - \nabla^\perp_{(\lambda}
  \,C_{\mu\underline{\rho}\nu)\sigma} u^{\rho}
  u^{\sigma}\,,
\end{align}
\end{subequations}
where $C_{\mu\nu\rho\sigma}$ stands for the Weyl tensor
\begin{equation}\label{eq:Weyl}
C_{\mu\nu\rho\sigma} = R_{\mu\nu\rho\sigma} - \Bigl(
  g_{\mu [ \rho} R_{\sigma ] \nu} - g_{\nu [ \rho} R_{\sigma ] \mu}
\Bigr) + \frac{1}{3}g_{\mu [ \rho}g_{\sigma ] \nu} R\,,
\end{equation}
and where we have used expressions for the original Weyl tidal moments in
which the STF operators have been removed or replaced by mere symmetrizations,
thanks to the properties of the Weyl tensor and the covariant derivative. To
start with,
we notice that, as one
can check, the Riemann and Weyl definitions of the current-type quadrupole
coincide, \textit{i.e.}, $H^{(C)}_{\mu\nu}=H^{(R)}_{\mu\nu}$. As a result, the
following discussion will, in fact, only concern the mass-type moments. From Eqs.
\eqref{eq:tidalRC}, we then get the following relations:
\begin{subequations}\label{eq:diffWeylRiemm}
\begin{align}
\left(G_{\mu \nu}G^{\mu \nu}\right)^{(C)} &= \left(G_{\mu \nu}G^{\mu
  \nu}\right)^{(R)} - G^{\mu\nu}_{(R)} R_{\mu \nu} + (\text{double
  zero terms}) \,, \label{eq:G2} \\
\left(G_{\mu \nu \rho}G^{\mu \nu \rho}\right)^{(C)} &= \left(G_{\mu
  \nu \rho}G^{\mu \nu \rho}\right)^{(R)}
-G^{\lambda\mu\nu}_{(R)}\nabla_\lambda R_{\mu\nu}- \frac{2}{3} u^\mu
u^\nu \nabla^\kappa R_{\kappa \mu\lambda\nu} \Big[\nabla^\lambda
  R_{\rho\sigma} u^\rho u^\sigma + \frac{1}{3} \nabla^\lambda R \Big]+
(\text{double zero terms}) \, , \label{eq:G3}
\end{align}
\end{subequations}
where the ``double zero terms'' are terms that are quadratic in the
Ricci tensor or scalar.  Let us now prove that the actions $S^{(R)}$
and $S^{(C)}$ corresponding to~\cref{eq:Sm3} using respectively the
Riemann and Weyl definitions, lead to the same EoM.

The double zero terms are treated as follows. Varying their contributions to
the action, which have necessarily the general form
$\propto \int \dd^4 x \sqrt{-g} A^{\mu\nu\rho\sigma...} \nabla_{...}
R_{\mu\nu}\nabla_{...} R_{\rho\sigma}$, leads, after possible integrations by
part, to a sum of terms
$\propto \int \dd^4 x \sqrt{-g} [\nabla_{...}(A^{\mu\nu\rho\sigma...}
\nabla_{...}R_{\rho\sigma})+ \nabla_{...}(A^{\rho\sigma\mu\nu...}
\nabla_{...}R_{\rho\sigma})] \delta R_{\mu\nu}$, plus surface integrals at
infinity which vanish, since their integrands contain factors
$\nabla_{...} R_{\rho\sigma}$ that are identically zero in vacuum. The
remaining terms are then proportional to (the covariant derivatives of) the
Ricci tensor multiplied by $A^{\mu\nu\rho\sigma...}$. On the other hand,
$A^{\mu\nu\rho\sigma...}$ is itself a sum of the form
$\sum_{A} \delta^{(4)}(x-y_A) F_A^{\mu\nu\rho\sigma...}$ and the presence of
the Dirac distributions forces the evaluation of the Ricci tensor to take
place at one particle's location, \textit{e.g.}, at $\bm{x}=\bm{y}_A$, in the
sense of dimensional regularization. Moreover, by virtue of Einstein's
equations (reinstalling the particles' label),
$[R_{\mu\nu}]_A = 8\pi [(T_\text{pp})_{\mu\nu}-
(T_{\text{pp}})^{\lambda}_{~\lambda} g_{\mu\nu}/2]_A +
\mathcal{O}(\epsilon_\text{tidal})$, where $[(T_\text{pp})^{\mu\nu}]_A$
denotes the point-particle stress-energy tensor of our particle system at
point $A$
\begin{equation}\label{eq:Tpp}
\left[T^{\mu\nu}_\text{pp}\right]_A = \sum_B m_B \int \dd \tau_B u_B^\mu u_B^\nu
\frac{\delta^{(4)}[y_A(\tau_A)-y_B(\tau_B)]}{\sqrt{-g}}\, .
\end{equation}
If
$A\neq B$, then $\delta^{(4)}[y_{A}(\tau_{A})-y_{B}(\tau_{B})] = 0$
because the compact objects never collide in the PN regime. If $A=B$,
the Dirac distribution reduces to $\delta^{(4)}(0)$, which is precisely
zero in dimensional regularization, as the limit of $\int \dd^{d+1}k\,
\de^{2\pi \di \,0}=0$ when $d\to 3$. Hence $[T^{\mu\nu}_\text{pp}]_A$
vanishes as well, and so does the contribution of the double-zero
terms to the Euler-Lagrange equations for the point-like bodies.

However, terms that are linear in both the Riemann and the Ricci tensors (or
the Ricci scalar) in Eqs.~\eqref{eq:diffWeylRiemm} cannot be dealt with in the
same way as the double zeros. Instead, they may be treated by making an
appropriate infinitesimal change of variable on the original metric, say
$g^\text{original}_{\mu \nu} = g_{\mu \nu} + h_{\mu \nu}$, in the action
$S^{(C)}[g^\text{original}_{\mu \nu},\bm{y}_A]$. This naturally defines the
new action
$\tilde{S}[g_{\mu\nu},\bm{y}_A] = S^{(C)}[g^\text{original}_{\mu
  \nu}[g_{\rho\sigma},\bm{y}_B],\bm{y}_A]$, dynamically equivalent to
$S^{(C)}$ when regarded as a functional of the metric $g_{\mu\nu}$. At first
order in $h_{\mu\nu}$, it reads
\begin{equation}\label{eq:deltaS}
\tilde{S}[g_{\mu \nu},\bm{y}_A] = S^{(C)}[g_{\mu \nu},\bm{y}_A] -
\frac{1}{16 \pi} \int \dd^{4}x \sqrt{-g}\left(R^{\mu \nu}-\frac{1}{2}R
g^{\mu \nu} -8\pi \,T^{\mu \nu} \right)h_{\mu \nu} + \mathcal{O}\left(
h^{2} \right)\, .
\end{equation}
Now, we want $\tilde{S}[g_{\mu \nu},\bm{y}_A]$ to coincide with
$S^{(R)}[g_{\mu \nu},\bm{y}_A]$. By choosing conveniently $h_{\mu\nu}$, the
term $(R^{\mu \nu}-\tfrac{1}{2}R g^{\mu\nu})h_{\mu\nu}$ will cancel the terms
linear in the Ricci tensor or scalar entering Eqs. \eqref{eq:diffWeylRiemm}.
As for the term $\int 
\dd^4 x\sqrt{-g}\, T^{\mu\nu}h_{\mu\nu}$, it vanishes by
itself and can thus be ignored. Indeed, integrating the
Dirac deltas contained in the expression chosen for $h_{\mu\nu}$ (see below) yields a sum
on $A=1,2$ of $\propto [T^{\mu\nu}]_A=[T^{\mu\nu}_{\text{pp}}]_A +
\mathcal{O}(\epsilon_\text{tidal})$, which boils down to
$\mathcal{O}(\epsilon_\text{tidal})$ since $[T^{\mu\nu}_{\text{pp}}]_A
=0$, as explained above around~\cref{eq:Tpp}.

Let us examine more precisely how to construct a $h_{\mu\nu}$
suitable to absorb the Ricci-type terms in
\cref{eq:deltaS} that come from the difference
$\Delta G_{\mu\nu}=G^{(C)}_{\mu\nu}-G^{(R)}_{\mu\nu}$. The contribution induced by
this difference through the 
modification of the mass quadrupole invariant
$\Delta (G_{\mu\nu}G^{\mu\nu}) = 2 G^{(R)}_{\mu\nu} \Delta G^{\mu\nu}+
(\text{double zero terms})$ has the form
\begin{equation} \label{eq:formDeltaS}
\int \dd^4 x \,Z_{\mu\nu} R^{\mu\nu}=\int \dd^4 x
\Big(Z_{\mu\nu}-\frac{1}{2}Z\indices{^\lambda_\lambda} g_{\mu\nu}\Big)
\Big(R^{\mu\nu}-\frac{1}{2}R g^{\mu\nu}\Big)\, .
\end{equation}
It is to be canceled by the piece
of the integral in \cref{eq:deltaS} that is sourced by $\propto
(R^{\mu\nu}-\tfrac{1}{2}R g^{\mu\nu})$. An obvious choice guaranteeing such
cancellation is
$h^{(G_{\rho\sigma})}_{\mu\nu}=16\pi(Z_{\mu\nu}-\tfrac{1}{2}Z\indices{^\lambda_\lambda}
g_{\mu\nu})/\sqrt{-g}$. Possible extra terms linear (at least) in the Ricci
tensor or scalar merely add irrelevant double zeros to the
action. Those can be tuned to have
\begin{equation}
h^{(G_{\rho\sigma})}_{\mu \nu} = - 4 \pi \sum_A \mu_A^{(2)} \int
\dd\tau_A \,\left[G^{(R)}_{\mu \nu}\right]_{A}
\,\frac{\delta^{(4)}[x^{\mu}-y_A^{\mu}(\tau_A)]}{\sqrt{-g}}\,.
\end{equation}

Regarding the mass octupole, we use the same method as for the
mass quadrupole to construct some suitable $h_{\mu\nu}^{(G_{\rho
    \sigma \tau})}$, the only new feature being that $\Delta
(G_{\lambda\mu\nu}G^{\lambda\mu\nu})$ is now a space-time integral
with a source of the form $Z^{\lambda}_{~\mu\nu}\nabla_{\lambda}
R^{\mu\nu}$.  However, the structure~\eqref{eq:formDeltaS} is
straightforwardly recovered by integrating by part. We finally find
that, in the mass-octupolar sector, the equality $\tilde{S}^{(G_{\rho
    \sigma \tau})}=(S^{(R)})^{(G_{\rho \sigma \tau})}$ is achieved by
setting:
\begin{align}
 h^{(G_{\rho \sigma \tau})}_{\mu \nu} &= 4 \pi
 \sum_{A}\frac{\mu^{(3)}_{A}}{3} \int \dd\tau_{A}\nabla^{\lambda}
 \left[ \left(G^{(R)}_{\lambda\mu\nu} +\frac{2}{3} \nabla^\kappa
   R_{\kappa\rho\lambda\sigma} u^\rho u^\sigma \Big(u_\mu
   u_\nu + \frac{2}{3} g_{\mu\nu} \Big) \right)_{\!\!A}
   \!\!\frac{\delta^{(4)}[x-y_{A}(\tau_{A})]}{\sqrt{-g}} \right]\,.
\end{align}

\section{The tidal acceleration to NNL order}
\label{appendix:accNNL}

By varying the total generalized Fokker
Lagrangian~\eqref{eq:Lpp}--\eqref{eq:Ltidal} and replacing
iteratively the accelerations by the values provided by the EoM
consistently truncated at lower orders, we obtain the total
acceleration of body 1 as $a_1^i = (a_1^i)_\text{pp} +
(a_1^i)_\text{tidal}$, where the point-particle part can be found in
\textit{e.g.}~\cite{BlanchetLR}, and where
\begin{align}\label{eq:a1tidal}
m_1 (a_1^i)_\text{tidal} &= \frac{G^2}{r_{12}^7} \Bigg\{ n_{12}^{i} \Bigl(-9
 m_{2}^{2} \mu_1^{(2)} 
 - 9 m_{1}^{2} \mu_2^{(2)}\Bigr)
 + \frac{1}{c^{2}} \biggl\{n_{12}^{i} \biggl [m_{2}^{2} \mu_1^{(2)}
 \Bigl(-36 (n_{12}{} v_{1}{})^2
 + 72 (n_{12}{} v_{1}{}) (n_{12}{} v_{2}{})\nonumber\\
& - 18 v_{12}{}^{2}
 + 9 v_{1}{}^{2}\Bigr)
 + m_{1}^{2} \mu_2^{(2)} \Bigl(144 (n_{12}{} v_{1}{})^2
 - 288 (n_{12}{} v_{1}{}) (n_{12}{} v_{2}{})
 + 180 (n_{12}{} v_{2}{})^2
 -  \frac{81}{2} v_{12}{}^{2}
 + 9 v_{1}{}^{2}\Bigr)\nonumber\\
& + m_{2}^{2} \sigma_1^{(2)} \Bigl(-96 (n_{12}{} v_{12}{})^2
 - 48 v_{12}{}^{2}\Bigr)
 + m_{1}^{2} \sigma_2^{(2)} \Bigl(-96 (n_{12}{} v_{12}{})^2
 - 48 v_{12}{}^{2}\Bigr)
 + \frac{G m_{1}}{r_{12}} \Bigl(\frac{159}{2} m_{2}^{2} \mu_1^{(2)}\nonumber\\
& + 132 m_{1}^{2} \mu_2^{(2)}\Bigr)
 + \frac{G m_{2}}{r_{12}} \Bigl(99 m_{2}^{2} \mu_1^{(2)}
 + 84 m_{1}^{2} \mu_2^{(2)}\Bigr)\biggl]
 + v_{1}^{i} \biggl [m_{2}^{2} \mu_1^{(2)} \Bigl(54 (n_{12}{} v_{1}{})
 - 45 (n_{12}{} v_{2}{})\Bigr)\nonumber\\
& + 9 m_{1}^{2} \mu_2^{(2)} (n_{12}{} v_{1}{})
 + 144 m_{2}^{2} \sigma_1^{(2)} (n_{12}{} v_{12}{})
 + 144 m_{1}^{2} \sigma_2^{(2)} (n_{12}{} v_{12}{})\biggl]
 + v_{2}^{i} \biggl [m_{2}^{2} \mu_1^{(2)} \Bigl(-54 (n_{12}{} v_{1}{})\nonumber\\
& + 45 (n_{12}{} v_{2}{})\Bigr)
 - 9 m_{1}^{2} \mu_2^{(2)} (n_{12}{} v_{1}{})
 - 144 m_{2}^{2} \sigma_1^{(2)} (n_{12}{} v_{12}{})
 - 144 m_{1}^{2} \sigma_2^{(2)} (n_{12}{} v_{12}{})\biggl]\biggl\}\nonumber\\
& + \frac{1}{c^{4}} \Biggl[n_{12}^{i} \biggl\{m_{2}^{2}
  \mu_1^{(2)} \Bigl(135 (n_{12}{} v_{1}{})^4
 - 540 (n_{12}{} v_{1}{})^3 (n_{12}{} v_{2}{})
 + 990 (n_{12}{} v_{1}{})^2 (n_{12}{} v_{2}{})^2\nonumber\\
& - 900 (n_{12}{} v_{1}{}) (n_{12}{} v_{2}{})^3
 + 225 (n_{12}{} v_{2}{})^4
 + 72 (n_{12}{} v_{1}{}) (n_{12}{} v_{2}{}) (v_{1}{} v_{2}{})
 - 18 (v_{1}{} v_{2}{})^2
 - 126 (n_{12}{} v_{1}{})^2 v_{12}{}^{2}\nonumber\\
& + 324 (n_{12}{} v_{1}{}) (n_{12}{} v_{2}{}) v_{12}{}^{2}
 - 90 (n_{12}{} v_{2}{})^2 v_{12}{}^{2}
 - 36 (v_{1}{} v_{2}{}) v_{12}{}^{2}
 - 27 v_{12}{}^{4}
 - 72 (n_{12}{} v_{1}{}) (n_{12}{} v_{2}{}) v_{1}{}^{2}\nonumber\\
& + 36 (v_{1}{} v_{2}{}) v_{1}{}^{2}
 + 36 v_{12}{}^{2} v_{1}{}^{2}
 - 18 v_{1}{}^{4}\Bigr)
 + m_{1}^{2} \mu_2^{(2)} \Bigl(-3855 (n_{12}{} v_{1}{})^4
 + 15420 (n_{12}{} v_{1}{})^3 (n_{12}{} v_{2}{})\nonumber\\
& - 23850 (n_{12}{} v_{1}{})^2 (n_{12}{} v_{2}{})^2
 + 16860 (n_{12}{} v_{1}{}) (n_{12}{} v_{2}{})^3
 - 4665 (n_{12}{} v_{2}{})^4
 - 288 (n_{12}{} v_{1}{}) (n_{12}{} v_{2}{}) (v_{1}{} v_{2}{})\nonumber\\
& + 360 (n_{12}{} v_{2}{})^2 (v_{1}{} v_{2}{})
 -  \frac{81}{2} (v_{1}{} v_{2}{})^2
 + 2598 (n_{12}{} v_{1}{})^2 v_{12}{}^{2}
 - 5484 (n_{12}{} v_{1}{}) (n_{12}{} v_{2}{}) v_{12}{}^{2}\nonumber\\
& + 3084 (n_{12}{} v_{2}{})^2 v_{12}{}^{2}
 - 81 (v_{1}{} v_{2}{}) v_{12}{}^{2}
 -  \frac{1923}{8} v_{12}{}^{4}
 + 288 (n_{12}{} v_{1}{}) (n_{12}{} v_{2}{}) v_{1}{}^{2}
 - 360 (n_{12}{} v_{2}{})^2 v_{1}{}^{2}\nonumber\\
& + 81 (v_{1}{} v_{2}{}) v_{1}{}^{2}
 + 81 v_{12}{}^{2} v_{1}{}^{2}
 -  \frac{81}{2} v_{1}{}^{4}\Bigr)
 + m_{2}^{2} \sigma_1^{(2)} \Bigl(840 (n_{12}{} v_{12}{})^4
 - 960 (n_{12}{} v_{12}{})^3 (n_{12}{} v_{1}{})\nonumber\\
& + 480 (n_{12}{} v_{12}{})^2 (n_{12}{} v_{1}{})^2
 - 192 (n_{12}{} v_{12}{})^2 (v_{1}{} v_{2}{})
 + 192 (n_{12}{} v_{12}{}) (n_{12}{} v_{1}{}) (v_{1}{} v_{2}{})
 - 48 (v_{1}{} v_{2}{})^2\nonumber\\
& - 336 (n_{12}{} v_{12}{})^2 v_{12}{}^{2}
 - 192 (n_{12}{} v_{12}{}) (n_{12}{} v_{1}{}) v_{12}{}^{2}
 + 192 (n_{12}{} v_{1}{})^2 v_{12}{}^{2}
 - 96 (v_{1}{} v_{2}{}) v_{12}{}^{2}
 - 72 v_{12}{}^{4}\nonumber\\
& + 192 (n_{12}{} v_{12}{})^2 v_{1}{}^{2}
 - 192 (n_{12}{} v_{12}{}) (n_{12}{} v_{1}{}) v_{1}{}^{2}
 + 96 (v_{1}{} v_{2}{}) v_{1}{}^{2}
 + 96 v_{12}{}^{2} v_{1}{}^{2}
 - 48 v_{1}{}^{4}\Bigr)\nonumber\\
& + m_{1}^{2} \sigma_2^{(2)} \Bigl(1000 (n_{12}{} v_{12}{})^4
 - 960 (n_{12}{} v_{12}{})^3 (n_{12}{} v_{1}{})
 + 480 (n_{12}{} v_{12}{})^2 (n_{12}{} v_{1}{})^2
 - 192 (n_{12}{} v_{12}{})^2 (v_{1}{} v_{2}{})\nonumber\\
& + 192 (n_{12}{} v_{12}{}) (n_{12}{} v_{1}{}) (v_{1}{} v_{2}{})
 - 48 (v_{1}{} v_{2}{})^2
 + 64 (n_{12}{} v_{12}{})^2 v_{12}{}^{2}
 - 192 (n_{12}{} v_{12}{}) (n_{12}{} v_{1}{}) v_{12}{}^{2}\nonumber\\
& + 192 (n_{12}{} v_{1}{})^2 v_{12}{}^{2}
 - 96 (v_{1}{} v_{2}{}) v_{12}{}^{2}
 - 128 v_{12}{}^{4}
 + 192 (n_{12}{} v_{12}{})^2 v_{1}{}^{2}
 - 192 (n_{12}{} v_{12}{}) (n_{12}{} v_{1}{}) v_{1}{}^{2}\nonumber\\
& + 96 (v_{1}{} v_{2}{}) v_{1}{}^{2}
 + 96 v_{12}{}^{2} v_{1}{}^{2}
 - 48 v_{1}{}^{4}\Bigr)
 + \frac{G m_{1}}{r_{12}} \biggl [m_{2}^{2} \mu_1^{(2)}
   \Bigl(\frac{7215}{8} (n_{12}{} v_{1}{})^2
 -  \frac{7431}{4} (n_{12}{} v_{1}{}) (n_{12}{} v_{2}{})\nonumber\\
& + \frac{4461}{8} (n_{12}{} v_{2}{})^2
 -  \frac{285}{8} v_{12}{}^{2}
 -  \frac{159}{2} v_{1}{}^{2}\Bigr)
 + m_{1}^{2} \mu_2^{(2)} \Bigl(- \frac{15717}{8} (n_{12}{} v_{1}{})^2
 + \frac{16581}{4} (n_{12}{} v_{1}{}) (n_{12}{} v_{2}{})\nonumber\\
& -  \frac{22521}{8} (n_{12}{} v_{2}{})^2
 + \frac{4597}{8} v_{12}{}^{2}
 - 132 v_{1}{}^{2}\Bigr)
 + m_{2}^{2} \sigma_1^{(2)} \Bigl(656 (n_{12}{} v_{12}{})^2
 - 144 (n_{12}{} v_{12}{}) (n_{12}{} v_{1}{})\nonumber\\
& + 200 v_{12}{}^{2}\Bigr)
 + m_{1}^{2} \sigma_2^{(2)} \Bigl(1124 (n_{12}{} v_{12}{})^2
 - 144 (n_{12}{} v_{12}{}) (n_{12}{} v_{1}{})
 + 436 v_{12}{}^{2}\Bigr)\biggl]\nonumber\\
& + \frac{G m_{2}}{r_{12}} \biggl [m_{2}^{2} \mu_1^{(2)} \Bigl(252 (n_{12}{} v_{1}{})^2
 - 504 (n_{12}{} v_{1}{}) (n_{12}{} v_{2}{})
 -  \frac{387}{2} (n_{12}{} v_{2}{})^2
 + 162 v_{12}{}^{2}
 - 99 v_{1}{}^{2}\Bigr)\nonumber\\
& + m_{1}^{2} \mu_2^{(2)} \Bigl(-2568 (n_{12}{} v_{1}{})^2
 + 5136 (n_{12}{} v_{1}{}) (n_{12}{} v_{2}{})
 - 2946 (n_{12}{} v_{2}{})^2
 + 426 v_{12}{}^{2}
 - 84 v_{1}{}^{2}\Bigr)\nonumber\\
& + m_{2}^{2} \sigma_1^{(2)} \Bigl(672 (n_{12}{} v_{12}{})^2
 + 336 v_{12}{}^{2}\Bigr)
 + m_{1}^{2} \sigma_2^{(2)} \Bigl(592 (n_{12}{} v_{12}{})^2
 + 192 v_{12}{}^{2}\Bigr)\biggl]
 + \frac{G^2 m_{1}^2}{r_{12}^2} \Bigl(- \frac{2145}{7} m_{2}^{2} \mu_1^{(2)}\nonumber\\
& - 1008 m_{1}^{2} \mu_2^{(2)}\Bigr)
 + \frac{G^2 m_{1} m_{2}}{r_{12}^2} \Bigl(- \frac{2581}{2} m_{2}^{2} \mu_1^{(2)}
 - 1805 m_{1}^{2} \mu_2^{(2)}\Bigr)
 + \frac{G^2 m_{2}^2}{r_{12}^2} \Bigl(-576 m_{2}^{2} \mu_1^{(2)}\nonumber\\
& -  \frac{6705}{14} m_{1}^{2} \mu_2^{(2)}\Bigr)\biggl\}
 + v_{1}^{i} \biggl\{m_{2}^{2} \mu_1^{(2)} \Bigl(-144 (n_{12}{} v_{1}{})^3
 + 468 (n_{12}{} v_{1}{})^2 (n_{12}{} v_{2}{})
 - 720 (n_{12}{} v_{1}{}) (n_{12}{} v_{2}{})^2\nonumber\\
& + 360 (n_{12}{} v_{2}{})^3
 - 342 (n_{12}{} v_{1}{}) (v_{1}{} v_{2}{})
 + 360 (n_{12}{} v_{2}{}) (v_{1}{} v_{2}{})
 + 144 (n_{12}{} v_{1}{}) v_{1}{}^{2}
 - 135 (n_{12}{} v_{2}{}) v_{1}{}^{2}\nonumber\\
& + 198 (n_{12}{} v_{1}{}) v_{2}{}^{2}
 - 225 (n_{12}{} v_{2}{}) v_{2}{}^{2}\Bigr)
 + m_{1}^{2} \mu_2^{(2)} \Bigl(1248 (n_{12}{} v_{1}{})^3
 - 3888 (n_{12}{} v_{1}{})^2 (n_{12}{} v_{2}{})\nonumber\\
& + 3996 (n_{12}{} v_{1}{}) (n_{12}{} v_{2}{})^2
 - 1392 (n_{12}{} v_{2}{})^3
 + 9 (n_{12}{} v_{1}{}) (v_{1}{} v_{2}{})
 -  \frac{903}{2} (n_{12}{} v_{1}{}) v_{12}{}^{2}
 + 492 (n_{12}{} v_{2}{}) v_{12}{}^{2}\nonumber\\
& - 9 (n_{12}{} v_{1}{}) v_{1}{}^{2}\Bigr)
 + m_{2}^{2} \sigma_1^{(2)} \Bigl(-1056 (n_{12}{} v_{12}{})^3
 + 1248 (n_{12}{} v_{12}{})^2 (n_{12}{} v_{1}{})
 - 576 (n_{12}{} v_{12}{}) (n_{12}{} v_{1}{})^2\nonumber\\
& - 960 (n_{12}{} v_{12}{}) (v_{1}{} v_{2}{})
 + 48 (n_{12}{} v_{1}{}) (v_{1}{} v_{2}{})
 + 336 (n_{12}{} v_{12}{}) v_{1}{}^{2}
 + 48 (n_{12}{} v_{1}{}) v_{1}{}^{2}
 + 624 (n_{12}{} v_{12}{}) v_{2}{}^{2}\nonumber\\
& - 96 (n_{12}{} v_{1}{}) v_{2}{}^{2}\Bigr)
 + m_{1}^{2} \sigma_2^{(2)} \Bigl(-1664 (n_{12}{} v_{12}{})^3
 + 1248 (n_{12}{} v_{12}{})^2 (n_{12}{} v_{1}{})
 - 576 (n_{12}{} v_{12}{}) (n_{12}{} v_{1}{})^2\nonumber\\
& - 1168 (n_{12}{} v_{12}{}) (v_{1}{} v_{2}{})
 + 48 (n_{12}{} v_{1}{}) (v_{1}{} v_{2}{})
 + 440 (n_{12}{} v_{12}{}) v_{1}{}^{2}
 + 48 (n_{12}{} v_{1}{}) v_{1}{}^{2}
 + 728 (n_{12}{} v_{12}{}) v_{2}{}^{2}\nonumber\\
& - 96 (n_{12}{} v_{1}{}) v_{2}{}^{2}\Bigr)
  + \frac{G m_{1}}{r_{12}} \biggl [m_{2}^{2} \mu_1^{(2)}
  \Bigl(- \frac{1209}{4} (n_{12}{} v_{1}{})
 + \frac{1179}{4} (n_{12}{} v_{2}{})\Bigr)
 + m_{1}^{2} \mu_2^{(2)} \Bigl(\frac{241}{4} (n_{12}{} v_{1}{})\nonumber\\
& -  \frac{661}{4} (n_{12}{} v_{2}{})\Bigr)
 + m_{2}^{2} \sigma_1^{(2)} \Bigl(-712 (n_{12}{} v_{1}{})
 + 856 (n_{12}{} v_{2}{})\Bigr)
 + m_{1}^{2} \sigma_2^{(2)} \Bigl(-1416 (n_{12}{} v_{1}{})
 + 1560 (n_{12}{} v_{2}{})\Bigr)\biggl]\nonumber\\
& + \frac{G m_{2}}{r_{12}} \biggl [m_{2}^{2} \mu_1^{(2)} \Bigl(-378 (n_{12}{} v_{1}{})
 + 279 (n_{12}{} v_{2}{})\Bigr)
 + m_{1}^{2} \mu_2^{(2)} \Bigl(714 (n_{12}{} v_{1}{})
 - 798 (n_{12}{} v_{2}{})\Bigr)\nonumber\\
& - 1008 m_{2}^{2} \sigma_1^{(2)} (n_{12}{} v_{12}{})
 - 784 m_{1}^{2} \sigma_2^{(2)} (n_{12}{} v_{12}{})\biggl]\biggl\}
 + v_{2}^{i} \biggl\{m_{2}^{2} \mu_1^{(2)} \Bigl(144 (n_{12}{} v_{1}{})^3
 - 468 (n_{12}{} v_{1}{})^2 (n_{12}{} v_{2}{})\nonumber\\
& + 720 (n_{12}{} v_{1}{}) (n_{12}{} v_{2}{})^2
 - 360 (n_{12}{} v_{2}{})^3
 + 342 (n_{12}{} v_{1}{}) (v_{1}{} v_{2}{})
 - 360 (n_{12}{} v_{2}{}) (v_{1}{} v_{2}{})
 - 144 (n_{12}{} v_{1}{}) v_{1}{}^{2}\nonumber\\
& + 135 (n_{12}{} v_{2}{}) v_{1}{}^{2}
 - 198 (n_{12}{} v_{1}{}) v_{2}{}^{2}
 + 225 (n_{12}{} v_{2}{}) v_{2}{}^{2}\Bigr)
 + m_{1}^{2} \mu_2^{(2)} \Bigl(-1248 (n_{12}{} v_{1}{})^3\nonumber\\
& + 3888 (n_{12}{} v_{1}{})^2 (n_{12}{} v_{2}{})
 - 3996 (n_{12}{} v_{1}{}) (n_{12}{} v_{2}{})^2
 + 1392 (n_{12}{} v_{2}{})^3
 - 9 (n_{12}{} v_{1}{}) (v_{1}{} v_{2}{})
 + \frac{903}{2} (n_{12}{} v_{1}{}) v_{12}{}^{2}\nonumber\\
& - 492 (n_{12}{} v_{2}{}) v_{12}{}^{2}
 + 9 (n_{12}{} v_{1}{}) v_{1}{}^{2}\Bigr)
 + m_{2}^{2} \sigma_1^{(2)} \Bigl(1056 (n_{12}{} v_{12}{})^3
 - 1248 (n_{12}{} v_{12}{})^2 (n_{12}{} v_{1}{})\nonumber\\
& + 576 (n_{12}{} v_{12}{}) (n_{12}{} v_{1}{})^2
 + 960 (n_{12}{} v_{12}{}) (v_{1}{} v_{2}{})
 - 48 (n_{12}{} v_{1}{}) (v_{1}{} v_{2}{})
 - 336 (n_{12}{} v_{12}{}) v_{1}{}^{2}
 - 48 (n_{12}{} v_{1}{}) v_{1}{}^{2}\nonumber\\
& - 624 (n_{12}{} v_{12}{}) v_{2}{}^{2}
 + 96 (n_{12}{} v_{1}{}) v_{2}{}^{2}\Bigr)
 + m_{1}^{2} \sigma_2^{(2)} \Bigl(1664 (n_{12}{} v_{12}{})^3
 - 1248 (n_{12}{} v_{12}{})^2 (n_{12}{} v_{1}{})\nonumber\\
& + 576 (n_{12}{} v_{12}{}) (n_{12}{} v_{1}{})^2
 + 1168 (n_{12}{} v_{12}{}) (v_{1}{} v_{2}{})
 - 48 (n_{12}{} v_{1}{}) (v_{1}{} v_{2}{})
 - 440 (n_{12}{} v_{12}{}) v_{1}{}^{2}
 - 48 (n_{12}{} v_{1}{}) v_{1}{}^{2}\nonumber\\
& - 728 (n_{12}{} v_{12}{}) v_{2}{}^{2}
 + 96 (n_{12}{} v_{1}{}) v_{2}{}^{2}\Bigr)
 + \frac{G m_{1}}{r_{12}} \biggl [m_{2}^{2}
 \mu_1^{(2)} \Bigl(\frac{1209}{4} (n_{12}{} v_{1}{})
 -  \frac{1179}{4} (n_{12}{} v_{2}{})\Bigr)\nonumber\\
& + m_{1}^{2} \mu_2^{(2)} \Bigl(- \frac{241}{4} (n_{12}{} v_{1}{})
 + \frac{661}{4} (n_{12}{} v_{2}{})\Bigr)
 + m_{2}^{2} \sigma_1^{(2)} \Bigl(712 (n_{12}{} v_{1}{})
 - 856 (n_{12}{} v_{2}{})\Bigr)\nonumber\\
& + m_{1}^{2} \sigma_2^{(2)} \Bigl(1416 (n_{12}{} v_{1}{})
 - 1560 (n_{12}{} v_{2}{})\Bigr)\biggl]
 + \frac{G m_{2}}{r_{12}} \biggl [m_{2}^{2} \mu_1^{(2)} \Bigl(378 (n_{12}{} v_{1}{})
 - 279 (n_{12}{} v_{2}{})\Bigr)\nonumber\\
& + m_{1}^{2} \mu_2^{(2)} \Bigl(-714 (n_{12}{} v_{1}{})
 + 798 (n_{12}{} v_{2}{})\Bigr)
 + 1008 m_{2}^{2} \sigma_1^{(2)} (n_{12}{} v_{12}{})
 + 784 m_{1}^{2} \sigma_2^{(2)} (n_{12}{} v_{12}{})\biggl]\biggl\}\Biggl]\nonumber\\
& + \frac{1}{r_{12}^{2}} n_{12}^{i} \Bigl(-60 m_{2}^{2} \mu_1^{(3)}
  - 60 m_{1}^{2} \mu_2^{(3)}\Bigr)\Bigg\}
  + \mathcal{O}\left( \frac{\epsilon_\text{tidal}}{c^{6}}
  \right)\,.
\end{align}
The tidal part of the relative acceleration in the CoM frame, deriving
from the CoM Lagrangian whose tidal part is shown
in~\eqref{eq:LCoMtidal}, reads
\begin{align}\label{eq:accCoM}
(a^{i})_\text{tidal} =& - 18 \frac{G^{2} m }{r^{7}}\mu_{+}^{(2)} n^{i}  \nonumber \\
& + \frac{1}{c^{2}} \left\{ \frac{G^{2} m}{r^{7}}
\left[ \left( \left(108 + 72\nu \right)\mu_{+}^{(2)} +
180 \Delta \, \mu_{-}^{(2)} -192 \sigma_{+}^{(2)} \right)\dot{r}^2 n^{i}  + 
\left( \left(-\frac{81}{2} -54\nu \right)\mu_{+}^{(2)}  - 
\frac{45}{2} \Delta \, \mu_{-}^{(2)} \right. \right. \right. \nonumber \\
& \left. \left. \left. -96 \sigma_{+}^{(2)} \right) v^{2} n^{i} + 
\left( \left(63-36\nu \right)\mu_{+}^{(2)} -45 \Delta \, \mu_{-}^{(2)} +
288\sigma_{+}^{(2)} \right)\dot{r} v^{i} \right] + \frac{G^{3} m^{2}}{r^{8}} 
\left[ \left(183+57\nu \right) \mu_{+}^{(2)} - 
15 \Delta \, \mu_{-}^{(2)} \right]n^{i} \right\} \nonumber \\
& + \frac{1}{c^{4}} \left\{ \frac{G^{2} m}{r^{7}}
\left[ \left( \left( -3720 -720 \nu +540 \nu^{2} \right)\mu_{+}^{(2)} + 
\left(-3990 +900 \nu \right) \Delta \, \mu_{-}^{(2)} + 
\left(880+960\nu \right)\sigma_{+}^{(2)} +
160 \Delta \, \sigma_{-}^{(2)}\right)\dot{r}^{4}n^{i} \right. \right. \nonumber \\
& \left. \left. + \left( \left( 2472 + 522 \nu -
288 \nu^{2} \right)\mu_{+}^{(2)} + 
\left(2724 -450 \nu \right) \Delta \, \mu_{-}^{(2)} -
272\sigma_{+}^{(2)} +400\Delta \, \sigma_{-}^{(2)} \right)
\dot{r}^{2}v^{2}n^{i} \right. \right. \nonumber \\
& \left. \left. + \left( \left( -\frac{1671}{8}-\frac{153}{2} \nu +
72 \nu^{2} \right)\mu_{+}^{(2)} + \left(-\frac{1527}{8} +
\frac{45}{2} \nu \right) \Delta \, \mu_{-}^{(2)} + 
\left(-104-96\nu \right)\sigma_{+}^{(2)} -
56 \Delta \, \sigma_{-}^{(2)}\right)v^{4}n^{i} \right. \right. \nonumber \\
& \left. \left. + \left( \left( 1104 +36 \nu -
144 \nu^{2} \right)\mu_{+}^{(2)} + 
\left(1392-180 \nu \right) \Delta \, \mu_{-}^{(2)} +
 \left(-1376-1536\nu \right)\sigma_{+}^{(2)} -
608 \Delta \, \sigma_{-}^{(2)}\right)\dot{r}^{3}v^{i}  \right. \right. \nonumber \\
& \left. \left. + \left( \left( -\frac{633}{2} +
63 \nu +36 \nu^{2} \right)\mu_{+}^{(2)} + 
\left(-\frac{1209}{2}+45 \nu \right) \Delta \, \mu_{-}^{(2)} + 
\left(872+672\nu \right)\sigma_{+}^{(2)} +
104 \Delta \, \sigma_{-}^{(2)}\right)\dot{r}v^{2}v^{i} \right] \right. \nonumber \\
& \left. + \frac{G^{3} m^{2}}{r^{8}}
\left[ \left( \left( -2316+\frac{5391}{4} \nu +
\frac{549}{2}\nu^{2} \right)\mu_{+}^{(2)} + 
\left(-2820 +\frac{1665}{4} \nu \right) \Delta \, \mu_{-}^{(2)} + 
\left(1264+744\nu \right)\sigma_{+}^{(2)} -
80 \Delta \, \sigma_{-}^{(2)}\right) \dot{r}^{2}n^{i} \right. \right. \nonumber \\
& \left. \left. + \left( \left( 405+\frac{887}{2} \nu -
  27 \nu^{2} \right)\mu_{+}^{(2)} + \left(279- \frac{135}{2}\nu \right)
 \Delta\,\mu_{-}^{(2)} + \left(528+216\nu \right)\sigma_{+}^{(2)} -
 144\Delta \, \sigma_{-}^{(2)}\right) v^{2}n^{i} \right. \right. \nonumber \\ 
& \left. \left. + \left( \left( 336-832 \nu - 114 \nu^{2} \right)\mu_{+}^{(2)} +
 \left(1092-150\nu \right)\Delta \, \mu_{-}^{(2)} +
 \left(-1792-960\nu \right)\sigma_{+}^{(2)} +
 224\Delta \, \sigma_{-}^{(2)}\right) \dot{r}v^{i}\right] \right. \nonumber \\
 & \left. + \frac{G^{4} m^{3}}{r^{9}}\left[
   \left( -\frac{14769}{14} - \frac{8716}{7} \nu \right)\mu_{+}^{(2)} +
   \left(\frac{1359}{14} +90 \nu\right)
   \Delta  \, \mu_{-}^{(2)}\right]n^{i} \right\} -
   120 \frac{G^{2} m }{r^{9}}\mu_{+}^{(3)}n^{i} + \mathcal{O}\left(
  \frac{\epsilon_\text{tidal}}{c^{6}} \right)\,.
\end{align}

\bibliography{ListeRef_HFB19.bib}

\end{document}